\documentclass[2pt]{iopart}

\usepackage{iopams}  
\usepackage{graphicx}
\usepackage{bm,upgreek}
\usepackage[breaklinks=true,colorlinks=true,linkcolor=blue,urlcolor=blue,citecolor=blue]{hyperref}
\usepackage{fixmath}
\usepackage{ulem}

\newcommand{\om}{\Omega}

\newcommand{\tno}{\theta_{n1}}
\newcommand{\tnt}{\theta_{n2}}

\newcommand{\beq}{\begin{eqnarray}}
\newcommand{\eeq}{\end{eqnarray}}
\newcommand{\bro}{\mathbf{r}}
\newcommand{\ba}{\mathbf{r}}
\newcommand{\bR}{\mathbf{R}}
\newcommand{\bV}{\mathbf{V}}
\newcommand{\bv}{\mathbf{v}}

\begin{document}

\title{Pairing, waltzing and scattering of chemotactic active colloids}

\author{Suropriya Saha$^{1,2}$}
\ead{Suropriya.saha@ds.mpg.de}

\author{Sriram Ramaswamy$^3$}
\ead{sriram@iisc.ac.in}

\author[cor1]{Ramin Golestanian$^{1,4}$}
\ead{ramin.golestanian@ds.mpg.de}
\vspace{.5cm}
\address{$^1$Max Planck Institute for Dynamics and Self-organisation, Am Fassberg 17, 307077 G\"ottingen, Germany} 
\address{$^2$Max Planck Institute for the Physics of Complex Systems, N\"othnitzer Stra{\ss}e 38, D-01187 Dresden, Germany}
\address{$^3$Centre for Condensed Matter Theory, Department of Physics, Indian Institute of Science, Bangalore 560 012, India}
\address{$^4$Rudolf Peierls Centre for Theoretical Physics, University of Oxford, Oxford OX1 3PU, United Kingdom}

\begin{abstract} 
An interacting pair of {chemotactic (anti-chemotactic)} active colloids, that can rotate their axes of self-propulsion to align {parallel (anti-parallel)} to a chemical gradient, shows dynamical behaviour that varies from bound states to scattering. The underlying two-body interactions are purely dynamical, non-central, non-reciprocal, and controlled by changing the catalytic activity and phoretic mobility. Mutually chemotactic colloids trap each other in a final state of fixed separation; the resulting `active dimer' translates. A second type of bound state is observed where the polar axes undergo periodic cycles leading to phase-synchronised circular motion around a common point. These bound states are formed depending on initial conditions and can unbind on increasing the speed of self propulsion. Mutually anti-chemotactic swimmers always scatter apart. We also classify the fixed points underlying the bound states, and the bifurcations leading to transitions from one type of bound state to another, for the case of a single swimmer in the presence of a localised source of solute.
\end{abstract}

\vspace{2pc}

%
\section{Introduction}

Gradients in the concentration of a solute along the surface of a particle in a fluid produce pressure differences, and thus a slip flow, parallel to the surface. If not anchored, the particle moves through the fluid in the direction opposite to the slip velocity. This phenomenon, known as diffusiophoresis \cite{Anderson-review,Frank_Prost}, finds dramatic application in the autonomous motility of active Janus colloids \cite{GLA2005, GLA2007,Kapral2007, Ebbens2012, Sen_Paxton}, that generate the required gradient of solute molecules themselves via a surface patch of catalyst that decomposes a species in the ambient medium. When placed in an \textit{externally imposed} gradient of solute, these artificial microswimmers rotate their intrinsic polarity, as defined by the catalyst patch, to point -- and thus to translate -- towards or away from regions of high concentration \cite{SRS, ExpChem1, BechingerExp}, emulating chemotaxis \cite{KS}.

Active colloids have emerged as a versatile constituent of synthetic active matter \cite{RMP_active}, as the surface profiles of catalyst concentration and solute-colloid interaction can be engineered to yield a variety of designed swimmer properties \cite{SRS,Soto}. Recent studies have explored their collective behaviour including pattern formation, motility induced phase separation, chemotactic collapse \cite{SRS, Wurger, Stark2, KapralTaxis, ThermoCollapse}, dynamic swarms \cite{JCcomet, ​ScatDefect, cates1, cates2, bocquet, Lowen_clusters, Lowen_capture} and oscillatory clustering \cite{Stark1, Stark2}. A study of \textit{two-particle} dynamics with a directory of possible interactions, crucial for a full understanding of the possible collective behaviour in these systems, is the subject of this paper. Earlier studies on pair interactions between motile particles include work on the role of hydrodynamic fields \cite{Ignacio_boundState,Goldstein_volvox,Yeomans-Dunkel,Yeomans-Pooley} and a general classification scheme \cite{ClassInter}. Our focus is on interactions mediated by the diffusing field of the reaction products.

\begin{figure}[h]
\begin{center}
\fl
                  \includegraphics[angle=0,width=3.7in]{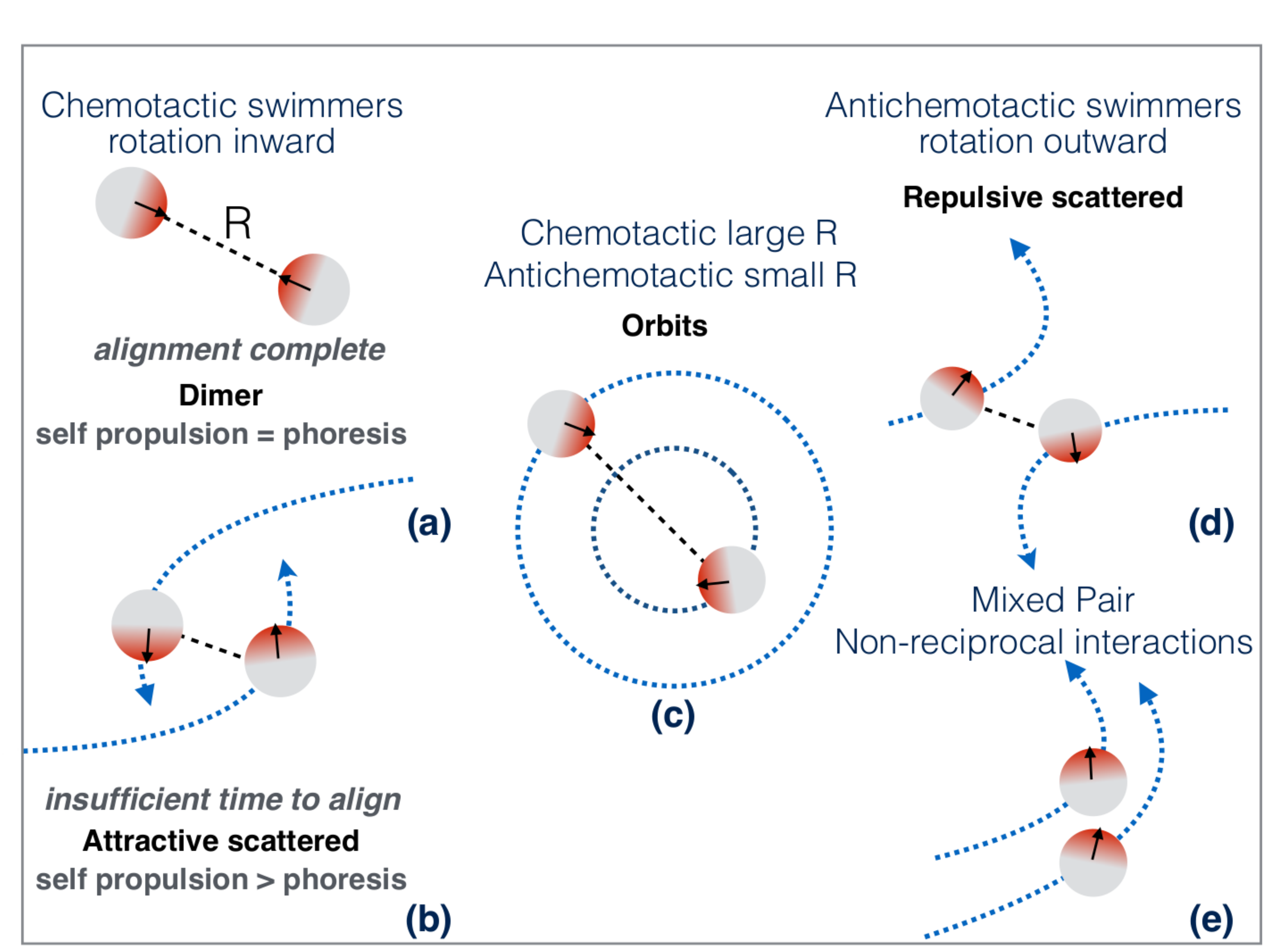}            
                  \fl \fl    
		    \caption{Schematics summarizing the final states. Sign of phoretic angular velocity that rotates the polar axis varies with the distance $R$ between swimmers. When it is positive (negative) at all $R$ for both swimmers, they are mutually chemotactic (anti-chemotactic) colloids. Self propulsion drives chemotactic (anti-chemotactic) swimmer towards (away from) one another. {A second contribution to centre of mass motion is from phoretic response to the chemical gradient}. (a) Alignment between a chemotactic pair leads to formation of a dimer by a cancellation of relative velocity. (b) Unsuccessful swimmers run away from one another. (c) Swimmers revolve around a common centre while maintaining a fixed angular separation between their axes. This happens when the sign of the phoretic angular velocity changes at small $R$. (d) Anti-chemotactic swimmers always scatter off one another. (e) A mixed pair shows explicit signatures of non-reciprocal interactions where one swimmer chases the other to form a bound pair.}
\label{Schematic}
        \end{center}
\end{figure}

Our pair of active colloids are spheres ($i=1,2$) of radius $\sigma$ characterised by a catalytic coating of surface density $A^{(i)}$ and a mobility $M^{(i)}$. The mobility $M^{(i)}$ is determined by the interaction potential between the surface of the sphere and the molecules of the product. Both vary from point to point on the surface of the sphere. We therefore decompose them into Legendre polynomials, assuming for simplicity that they are axisymmetric with a common polar axis:
 \begin{eqnarray}
A^{(i)}(\theta_i) =  a^{(i)} {\sum}_{\ell} \alpha^{(i)}_{\ell} P_{\ell} (\theta_i), \,  
M^{(i)}(\theta_i) &=& m^{(i)} {\sum}_{\ell} \mu^{(i)}_{\ell} P_{\ell} (\theta_i),
\label{source}
\end{eqnarray} 
where $a^{(i)}$ is the average rate of production of product molecules per unit area, $m^{(i)}$ sets the scale of the mobility and $\theta_i$ is the colatitude on the sphere. The essential phenomenology is conveyed by the simple case where we retain $\ell = 0,1,2$ for $\alpha^{(i)}_\ell$ and $\mu^{(i)}_\ell$. The surface patterns are represented by the sets of numbers $ \{\mu^{(i)}_{0},\mu^{(i)}_{1},\mu^{(i)}_{2}\}$ and $ \{\alpha^{(i)}_{0},\alpha^{(i)}_{1},\alpha^{(i)}_{2}\}$, with each such assignment defining a distinct swimmer design.  As shown in \cite{SRS}, an active colloid with $\mu^{(i)}_{\ell} \neq 0$ for $\ell>0$, generates chemotactic angular velocity that can turn its polar axis to align with a gradient. Parallel and antiparallel alignment are respectively called chemotactic and anti-chemotactic.
\begin{figure}
\begin{center}
                  \includegraphics[angle=0,width=4.2in]{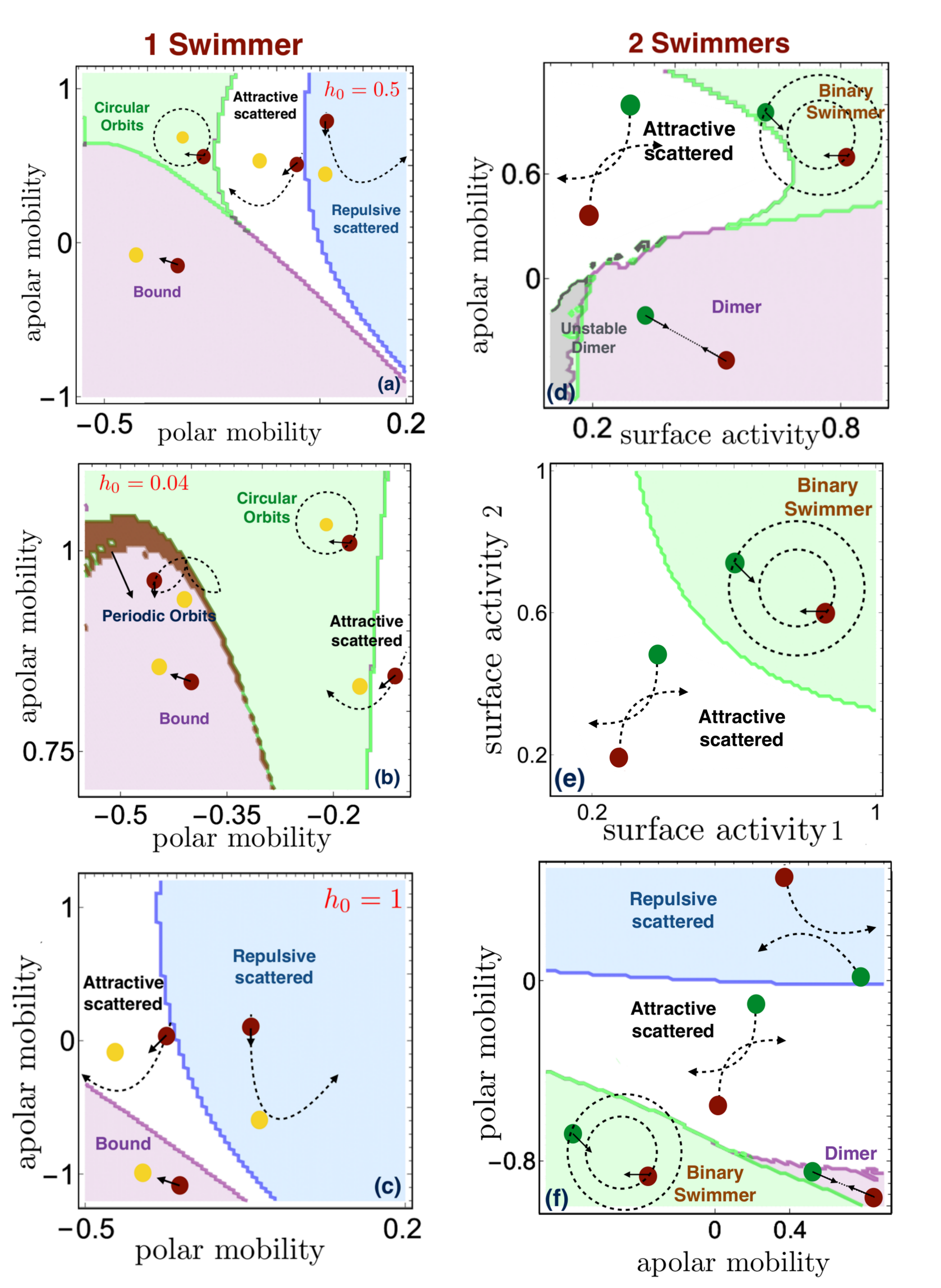}                
		    \caption{{2D state diagrams categorizing steady state dynamics are constructed by varying paired combinations of polar mobility $\mu_{1}^{(i)}$, apolar mobility $\mu_{2}^{(i)}$ and total surface activity $a^{(i)}$, keeping other parameters and initial conditions fixed (for details, see Appendix A). Figs. (a)-(c) on the left show the trapped, orbiting and scattered states observed when a swimmer interacts with a source for three values of $h_0 = a^{(1)}/a^{(2)}$. In figs. (a) and (c) the source produces an isotropic districution of product while in fig. (b) it is anisotropic. Increasing $h_0$, equivalent to increasing the speed of self-propulsion, tends to eliminate bound states by causing scattering as can be seen by comparing figs. (a) and (c) which differ in $h_0$ only. Fig. (d)-(f) on the right show bound states for two mobile swimmers showing active dimers, binary-swimmers and scattering.}}		    	    
\label{StateDiagram}
        \end{center}
\end{figure}

Here is a summary of our results. We consider two distinct cases -- a swimmer
interacting with a fixed source of solute and two swimmers that are both free to move. (i) A colloid which is (positively) chemotactic at every separation from the source, $\mu_1^{(1)}<0, \mu_2^{(1)}<0$, aligns with the gradient and is trapped by the source when self-propulsion and phoretic repulsion balance.  (ii) A colloid that exhibits mixed response to a the source, chemotactic at large separation and antichemotactic when within a few particle radii from source, executes what appears to be periodic motion in closed orbits. (iii) Increasing the speed of the swimmer or increasing the impact parameter drives a transition from bound state to scattering. (iv) An anti-chemotactic colloid is always scattered. Turning to the case of two swimmers we find: (v) Two interacting swimmers can form a stationary or moving dimer stabilised by a balance of self propulsion and phoretic repulsion,  with a fixed distance between them.  (vi) Two swimmers, at least one of which has a mixed chemotactic response to the other, can form bound states where they revolve around a common point on closed phase synchronised orbits while their polar axes are locked at a finite inclination. (vii) Two {anti-chemotactic} swimmers always scatter off one another. The steady states have been  summarised in state diagrams (see fig. \ref{StateDiagram}) obtained by varying total surface activity $a^{(i)}$ and composition of the motility coat. Finally, we also show that the bound states are robust to the presence of thermal or other fluctuations, so that signatures of these states should be seen in experiments.

The article is organised as follows: In section 2 we present the equations of motion for the swimmers for a given separation and relative orientation and outline steps followed to obtain the dynamics. In section 3 we discuss the bound states obtained when one swimmer interacts with a source of solute. In section 4 we categorize the dynamics shown by two mobile swimmers. 

\section{Interacting active colloids}
Consider two swimmers in a fluid medium constrained to move in a plane that also contains their polar axes. Reactants are converted into products when they come in contact with the enzyme coated colloidal surface, resulting in a spatiotemporal distribution $\rho(\bro,t)$ of products. The product diffuses freely in the bulk of the fluid with diffusion coefficient $D$. {There is a normal flux of the product molecules on surfaces of the swimmers at every point $\ba_i$}
\beq
-D\nabla^2 \rho = 0,\,\, -D {\nabla}_{\perp} \rho(\ba_i,t)  = A^{(i)}(\ba_i).
\label{TwoC}
\eeq
The activity $a^{(i)}$, can depend on the availability of reactant locally in which case it can be approximated by the Michaelis-Menten \cite{MM} factor. In this work we do not consider this aspect and work with constant $a^{(i)}$.

\begin{figure}[h]
\begin{center}
                \includegraphics[angle=0,width=2in]{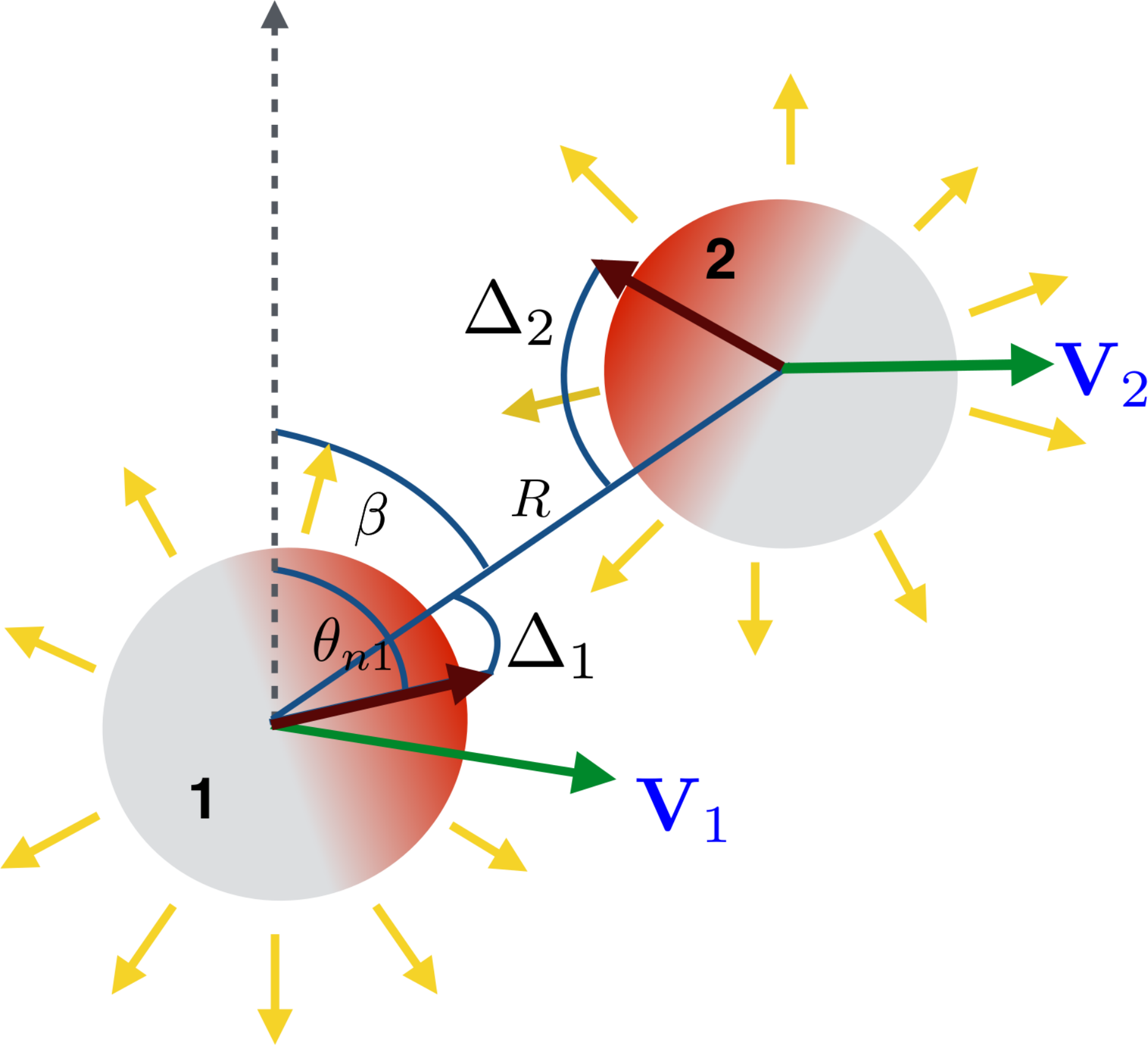}        
	    \caption{{Schematic showing the relative position of the two swimmers. Swimmer velocities point in directions different from the line joining their centres, leading to  non-central and non-reciprocal interactions. {By symmetry, equations of motion for colloid position and polarity can depend only on the relative orientations $\Delta_{1,2}$ and separation $R$}. Fixed points of this two-body system is determined by stationary values of the three relative parameters while actual motion occurs in the six dimensional space leading to interesting oscillatory states. }}
\label{TwoColloids}
     \end{center}
\end{figure}

Under conditions of vanishing Reynolds number, the mobility and chemical gradient lead to the establishment of a slip velocity $\bv_{i} = M^{(i)} \mathbf{\nabla}_\parallel \rho$. The slip velocity produces linear and angular velocities which for a sphere are given by the following surface integrals
\begin{eqnarray}
 \mathbf{V}_i =- \frac{1}{4 \pi }\int  \mbox{d}s \, \,  \bv_i (\ba_i), \,\,
\bm{\omega}_i =- \frac{3}{16 \pi \sigma_i } \int \mbox{d}s \,\, {\ba}_i \times \bv_{i} (\ba_i) .
\label{B2}
\end{eqnarray}
The linear velocity $\mathbf{V}_i$ subsumes both self propulsion and phoretic interactions, while $\bm{\omega_i}$ is the angular velocity of the swimmer. The sign of $\bm{\omega}_i$ determines the `chemotactic' response of  the swimmer i.e. whether it aligns to point up or down the gradient \cite{SRS}.  Equations of motion for the positions $\bR_{1,2}$ and orientations $\tno,\tnt$ of the swimmers in the lab frame are built by first solving eq. \ref{TwoC} for $\rho$. The solution for $\rho$ gives $\bv_i$ which is inserted into eq. \ref{B2} to calculate the velocities. Translation and rotation invariance of space restricts the dependence of these velocities on $R$ and the orientation of the polar axes, $\Delta_1$ and $\Delta_2$, measured with respect to the line joining the sphere centres (see fig. \ref{Schematic}). {In Appendix B we show that the field $\rho$ can be calculated to a desired degree of precision as a perturbation series in powers of $\sigma/R$, where $\sigma$ is the colloid size. The calculation uses a method of reflections, analogous to the method of images used in electrostatics. For simplicity, we construct the dynamics for just one set of reflections; corrections from subsequent reflections contribute to progressively larger powers in $\sigma/R$ and can be ignored.}  

The linear velocity of swimmer 1 in spherical polar coordinates in the frame of reference of swimmer 2 in terms of radial and angular velocities $V$ and $\Omega$ is, $\bV_1 = V \hat{R} + \Omega R \hat{\beta}_1$. $\hat{R}$ and $\hat{\beta}_1$ are unit vectors in the radial and tangential directions. We assume the swimmer size $\sigma$ to scale distances, while the velocities are scaled by $  a^{(1)} m^{(1)} /D$. The equations of motion for swimmer 1 in terms of the separation $R$  and the relative orientations $\Delta_1 = \theta_{n1} - \beta_1$ and $\Delta_2 = \theta_{n2} - \beta_1 -\pi$ are 
\beq
\fl
&& \dot{ \theta}_{n1} = \frac{a^{(1)} m^{(1)}}{ D \sigma} \omega \left( \Delta_1,\Delta_2, R \right), \\
&& \dot{R_1} = \frac{a^{(1)} m^{(1)}}{ D }  V \left(\Delta_1,\Delta_2, R \right), \\ 
&& \dot{\beta_1} = \frac{a^{(1)} m^{(1)}}{ D}  \Omega \left(\Delta_1,\Delta_2, R \right).
\label{veldef}
\eeq
The velocity of swimmer 1 is a linear superposition of contribution from different harmonics of the catalytic coat of swimmer 2, we first discuss the case when $\alpha_0^{(2)} =1$ and $\alpha^{(2)}_\ell = 0 $ for $\ell>0$. The angular velocity for the polar axis is
\beq
\omega(\Delta_1,\Delta_2,R)  &=& \frac{3 \sigma^2 \mu_1^{(1)}}{8 R^2} \sin \Delta_1 + \frac{3 \sigma^3\mu_2^{(1)}}{8 R^3} \sin 2\Delta_1. 
\label{expom} 
\eeq
At a given position, the polar axis rotates till it aligns parallel ($\Delta_1 = 0$) or anti-parallel ($\Delta_1=\pi$) with the local concentration gradient. The nature of the chemotactic response of the swimmer is determined by the sign of the factor $ \mu_1^{(1)} + \mu_2^{(1)}/ R $, for small deviations around alignment. When $\mu_1^{(1)}, \mu_2^{(1)}$ carry the same sign, the chemotactic response stays unchanged  at all distances from the source, being  chemotactic (anti-chemotactic) for negative (positive) sign. For $\mu_1^{(1)}<0$ and $\mu_2^{(1)}>0$, the response depends on radial distance and reverses when the swimmer approaches the source, leading to orbiting states which we will discuss in the next section.  The radial and angular velocities are 
\beq
\fl
V(\Delta_1,\Delta_2,R) &= & -\frac{\sigma^2 }{ R^2} \left( \mu_{0}^{(1)} - \frac{\mu_2^{(1)}}{20} \right)  + \frac{3  \sigma^2 \mu_2^{(1)} }{20 R^2} \cos 2\Delta_1  - \frac{ 2  \sigma^3 \mu_1^{(1)} }{3 R^3} \cos \Delta_1  \nonumber \\ 
\fl
&& -  \frac{h_0}{15} (5 \alpha_{1}^{(1)} \mu_{0}^{(1)} + 2 \alpha_{2}^{(1)} \mu_1^{(1)}-\alpha_{1}^{(1)} \mu_{2}^{(1)}) \cos \Delta_1 ,\label{expvelR}\\
\fl
\Omega(\Delta_1,\Delta_2, R) &=&  \frac{ \sigma^4 \mu_1^{(1)} }{3 R^4} \sin \Delta_1 + \frac{3  \sigma^3 \mu_2^{(1)} }{20 R^3} \sin 2\Delta_1 \nonumber \\  
&& -  \frac{h_0}{15} (5 \alpha_{1}^{(1)} \mu_{0}^{(1)} + 2 \alpha_{2}^{(1)} \mu_1^{(1)}-\alpha_{1}^{(1)} \mu_2^{(1)} ) \sin \Delta_1.
\label{expvelang}
\eeq
Here the dimensionless parameter $h_0= {a^{(1)}}/{a^{(2)}}$ is the ratio of the total catalytic activity of swimmers 1 and 2. In eqs. \ref{expvelR} and \ref{expvelang}, the terms proportional to $h_0$ arise from self propulsion and the rest are due to phoretic response to product field of swimmer 2. As swimmer 1 responds to the product field generated by itself and that produced by swimmer 2 with the same mobility, $h_0$ measures the relative strength of self propulsion and interaction. The higher order terms in $\omega$ and $\bV$ are always of the form of trigonometric functions of $p \Delta_1$, where $p$ is an integer. When the swimmers are sufficiently separated, the polarity relaxes before the position, as it can do so independently of the separation. Assuming this separation of timescale between the radial position and polarity, the mutual radial velocity can be obtained by substituting $\Delta_1 =0$ in eq. \ref{expvelR}. The non monotonic radial velocity, which is a consequence of varying mobility and self propulsion, affords the possibility of it vanishing at finite $R$ producing bound states. The most dominant contribution to the radial velocity is the $1/R^2$ term which causes a repulsion of the swimmer from the source for $\mu^{(1)}_0>0$. For $\mu^{(1)}_0>0$ and $\alpha^{(1)}_1<0$, self propulsion is directed opposite to phoretic repulsion thus leading to formation of a bound state. The details of these calculations are presented in the next section.   

Finally,  we write down a few of the terms in $\omega$ that are present when the anisotropy of chemical field produced by swimmer 2 is considered,  and present the full equations in the appendix. 
\beq
\omega  &\approx & \frac{3  \sigma^3 \alpha_{1}^{(2)} \mu_1^{(1)}}{32 R^3} \left[ \sin \Delta_1+ 3 \sin(\Delta_1 +  \Delta_2)  \right]  \nonumber \\
 &-& \frac{3 \sigma^4  \alpha_{2}^{(2)} \mu_1^{(1)}}{64 R^4} \left[  5 \sin (\Delta_1+2 \Delta_2) + \sin (\Delta_1-2 \Delta_2) + 2 \sin \Delta_1\right]. 
\label{anisovel}
\eeq
The colloid is a micron sized particle subject to fluctuations due to thermal as well as non-thermal sources, so that the dynamics should be supplemented with appropriate noises. We present a discussion on the effect of noise on the dynamics in a later section. In the next section we can discuss two possible bound states and explore swimmer designs that lead to one or the other. 

\section{Single active colloid in a product source}\label{SingleColloid}
We first consider the case when, $m^{(2)}=0$ and $m^{(1)} \neq 0$; swimmer 1 interacts with a tethered swimmer 2, that we now call source. {In this section, an isotropic source is one that is uniformly coated with catalysts, $\alpha_{\ell}^{(2)} = 0$ for $\ell >0$;  an anisotropic source has at least one nonzero $\alpha_{\ell}^{(2)}$ for $\ell>0$.}  

\subsection{Bound final state}\label{Trapped}
\begin{figure}[h]
\begin{center}
                  \includegraphics[angle=0,width=5.5in]{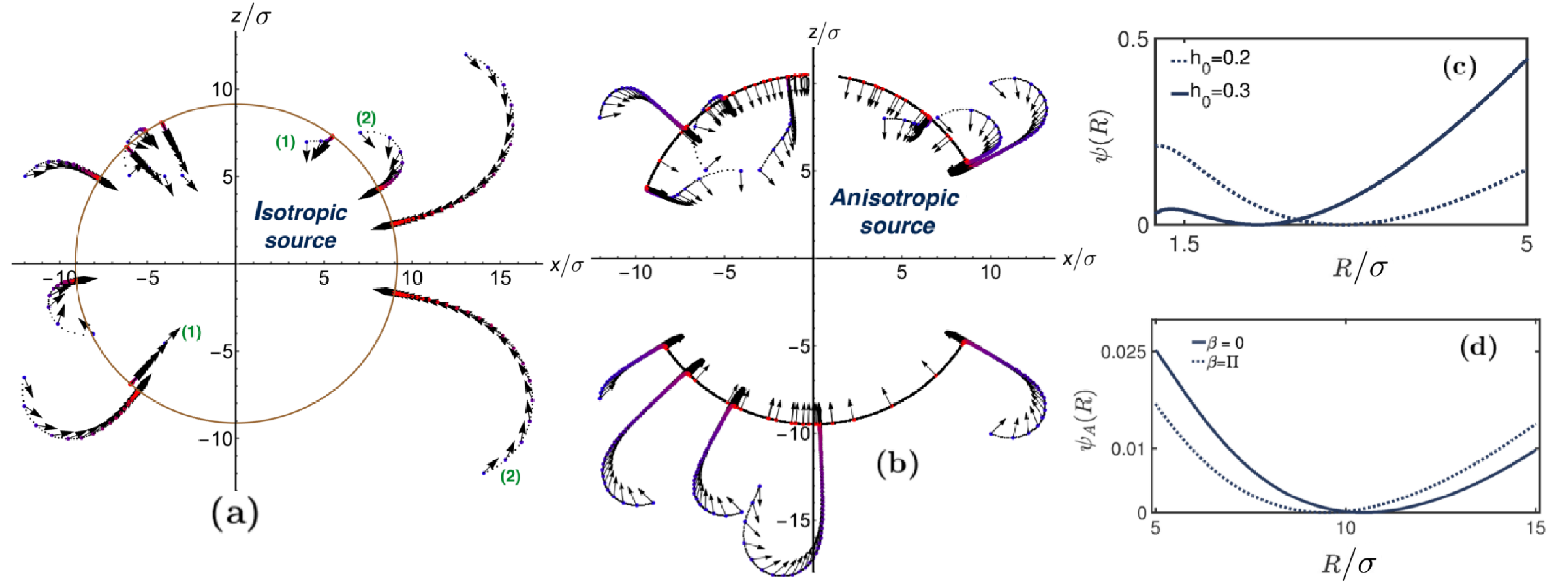}
		    \caption{{Typical paths traced by a swimmer near isotropic and anisotropic sources are shown in figs. (a) and (b) respectively. The coloured markers represent swimmer position, arrows represent polar axis. Marker colour changes from blue to red to with time. An isotropic source produces a radially symmetric effective potential plotted in fig. (c) as a function of $R$ for different choices of $h_0$. The swimmer is strongly confined for larger self propulsion. An anisotropic source traps the swimmer preferentially along its axis of symmetry producing an anisotropic potential plotted in fig. (d) for motion along $\beta =0,\pi$. The mobility and catalytic coat designs are $\{ 0.3,-0.5, 0.4 \} $ and $\{1, -1, 0\}$ respectively. The catalytic coat of source is $\{1,1,0\}$. $h_0 = 0.03$ for fig. (a), (b) and (d).}}
\label{Isotropic}
        \end{center}
\end{figure}
We first discuss a bound state formed by a swimmer with mobility $\mu_1^{(1)}<0$ and $\mu_2^{(1)}<0$. The swimmer can be trapped on a collection of points where both translational and angular velocities vanish. A swimmer that impinges on a source aligns with the local concentration gradient, provided it spends enough time in its vicinity. It is trapped if the radial velocity vanishes. For an isotropic source, the trapping surface is a circle of radius $R_0$ centred at the source. The final orientation is $\Delta_1 = 0$, so that $R_0$ satisfies the condition $V(0,0,R_0)=0$. Using eq. \ref{expvelR}, and ignoring terms of order higher than $\sigma^2/R_0^2$, we find  
\beq
R_0 \approx \sigma \sqrt{\frac{3}{h_0}} \left(- \alpha_1^{(1)} - \frac{2 \alpha_2^{(1)} \mu^{(1)}_1}{5 \mu^{(1)}_0  - \mu^{(1)}_2}  \right)^{-\frac{1}{2}}.
\eeq
Swimmers designed with parameters that allow a positive $R_0$ are trapped. $R_0$ depends on swimmer design and can be tuned by varying the ratio $\alpha^{(1)}_0/\alpha^{(1)}_2$, for example. Swimmers of distinct designs are thus trapped at different distances from the source. This provides a method for sorting the colloids experimentally. 

The motion of the swimmer in a chemical field can be equivalent to colloid motion in an externally applied attractive force field in a dissipative medium.  This {is} seen as follows: with the assumption that $\Delta_1$ has relaxed to its stationary value, the equation of motion of the colloid for deviations away from $R_0$, {denoted by $\delta R$} is
\beq
 \dot{\delta R}(t) = V(0,0, R=R_0 \pm \delta R) \equiv -\mathbf{\nabla} \psi.  
\eeq
The source produces a confining potential $\psi(R) = \int_{R_0}^R \mbox{d} R' V(0,0,R')$, when $\partial_R^2 \psi(R)>0$. An isotropic source produces a radially symmetric potential, see fig. \ref{Isotropic} (c). {Near an anisotropic source}, $\omega$ vanishes when $\tno - \tnt=\pi$, {thus creating} an anisotropic confining potential with {two traps along the polar axis of the source}. The swimmer gets trapped in one or the other depending on the initial conditions. To illustrate the strength of the confinement we plot the anisotropic potential produced along the source axis in fig. \ref{Isotropic} (d). {The region of state space where the bound state is formed is shown in pink in fig. \ref{StateDiagram} (a) - (c). Note a swimmer is trapped either when both $\mu^{(1)}_{1,2}<0$ or either one of them is both negative and sufficiently larger in magnitude than the other.} 

\subsection{Orbits}\label{Orbits}
\begin{figure}[h]
\begin{center}
     \includegraphics[angle=0,width=5in]{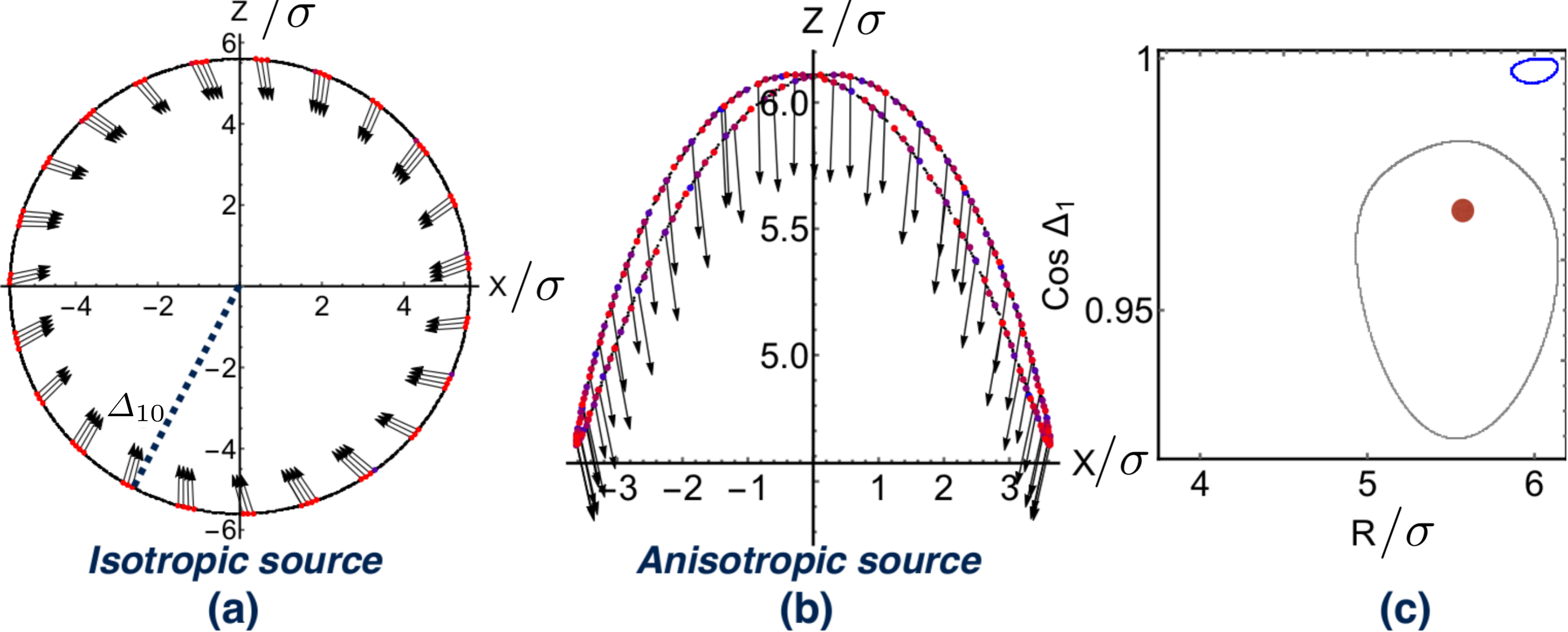}
	    \caption{{Swimmer orbits around the source tracing periodic circular orbits in fig. (a) which become non circular as the source is made an-isotropic in fig. (b). The parameters for the source are same as in Fig. \ref{Isotropic} and the parameters for $\{ \mu^{(1)}_{\ell}\}$ of the swimmer are $\{0.3,-0.5,1\}$. In the case of an anisotropic source, the fixed point exists in three dimensional space spanned by $\Delta_{1},\beta,R$. In (c) we show two dimensional projections in $\Delta_1-R$ plane for $h_0=0.04$ (blue line)and $h_0=0.06$ (gray line). The red point is the fixed point for an isotropic source for $h_0=0.04$.}}
\label{Orbits}
     \end{center}
\end{figure} 
A swimmer with $\mu_2^{(1)}>0$ and $\mu_1^{(1)}<0$, can execute persistent periodic orbits around the source. This particular choice of parameters ensures that the swimmer is chemotactic at large distance and anti-chemotactic near the source. Close to an isotropic source, it tries to align with the local gradient and becomes trapped in a limit cycle at a constant distance $R_0$ (indicated in green in fig. \ref{StateDiagram} (a) and (b)) while maintaining a constant angle $\Delta_{10}$ with the line joining its centre with the source. The angle $\Delta_{10}$ and radius $R_0$   are determined from conditions
\beq
 \omega(\Delta_{10},0,R_0)  - \om(\Delta_{10},0,R_0) = 0 , \,\,\,V(\Delta_{10},0,R_0)  = 0.
\label{orbit_condt}
\eeq
In an anisotropic chemical field, we find non-circular orbits of two distinct types -  trajectories that resemble a figure of 8 and do not enclose the source and non-circular orbits that do (see fig. \ref{Orbits} (b) and brown regions of state diagram in fig. \ref{StateDiagram} (b)). These oscillations can occur in this inertia-less regime because of the dynamics of the extra degree of freedom, namely, the polar axis. 
\newpage
\subsection{Effect of fluctuations on bound states}
\begin{figure}[h]
\begin{center}
                  \includegraphics[angle=0,width=4.2in]{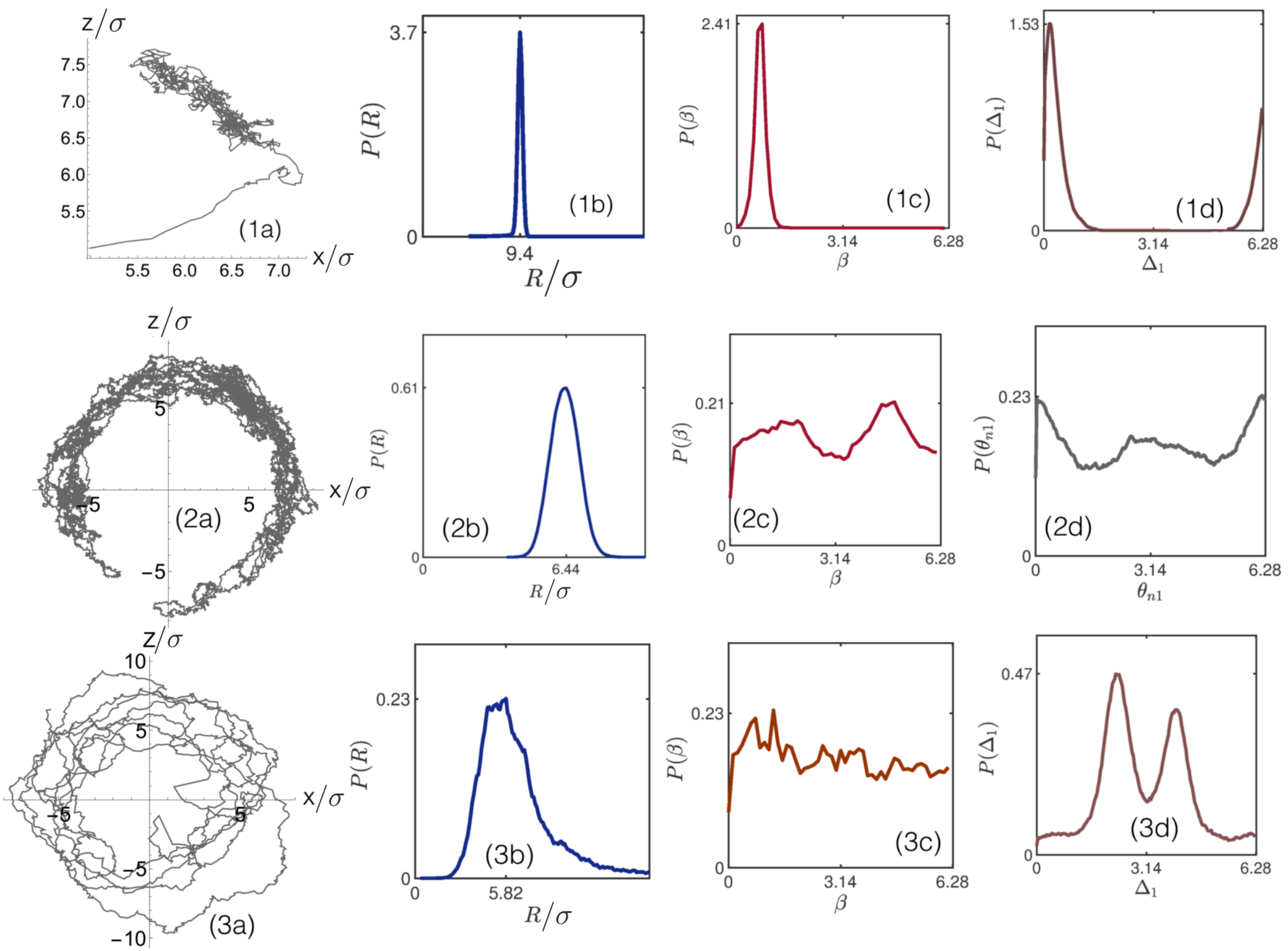}
		    \caption{{Figure showing the stability of the single bound states to fluctuations. The swimmer parameters in fig. (1a) and fig. (3a) are same the same as in fig. \ref{Isotropic} and while that in fig. (3a) is same as in fig. \ref{Orbits}. The stationary distribution of $R$, $\Delta_1$ and $\beta$ have been plotted to show that they peak at the fixed points for the noise-free case. These distributions should be measurable in experiments. The general nature of the distributions can be produced form experimental data and particular features like the bimodal distributions in figs. (2c), (2d) and (3d) would be signals of an-isotropic trapping and orbiting states.  }}
\label{Fluctuations}
        \end{center}
\end{figure}
To check the stability of the bound states to thermal fluctuations, we add Gaussian white noise terms are added to the dynamics in eqs. \ref{expom}, \ref{expvelR}, \ref{expvelang} with the strength chosen as appropriate for a free swimmer of Peclet number $15$. The noisy trajectories so obtained and distributions of $R$, $\Delta_1$ and $\beta$ are shown in fig. \ref{Fluctuations}. The values of $R \sigma^{-1}$ indicated on the axes are the mean values computed from the distribution and they match the deterministic values of $R_0$ shown in figs. \ref{Isotropic} and \ref{Orbits}.
 
\newpage
\subsection{Scattering}
\begin{figure}[h]
\begin{center}
                  \includegraphics[angle=0,width=3.5in]{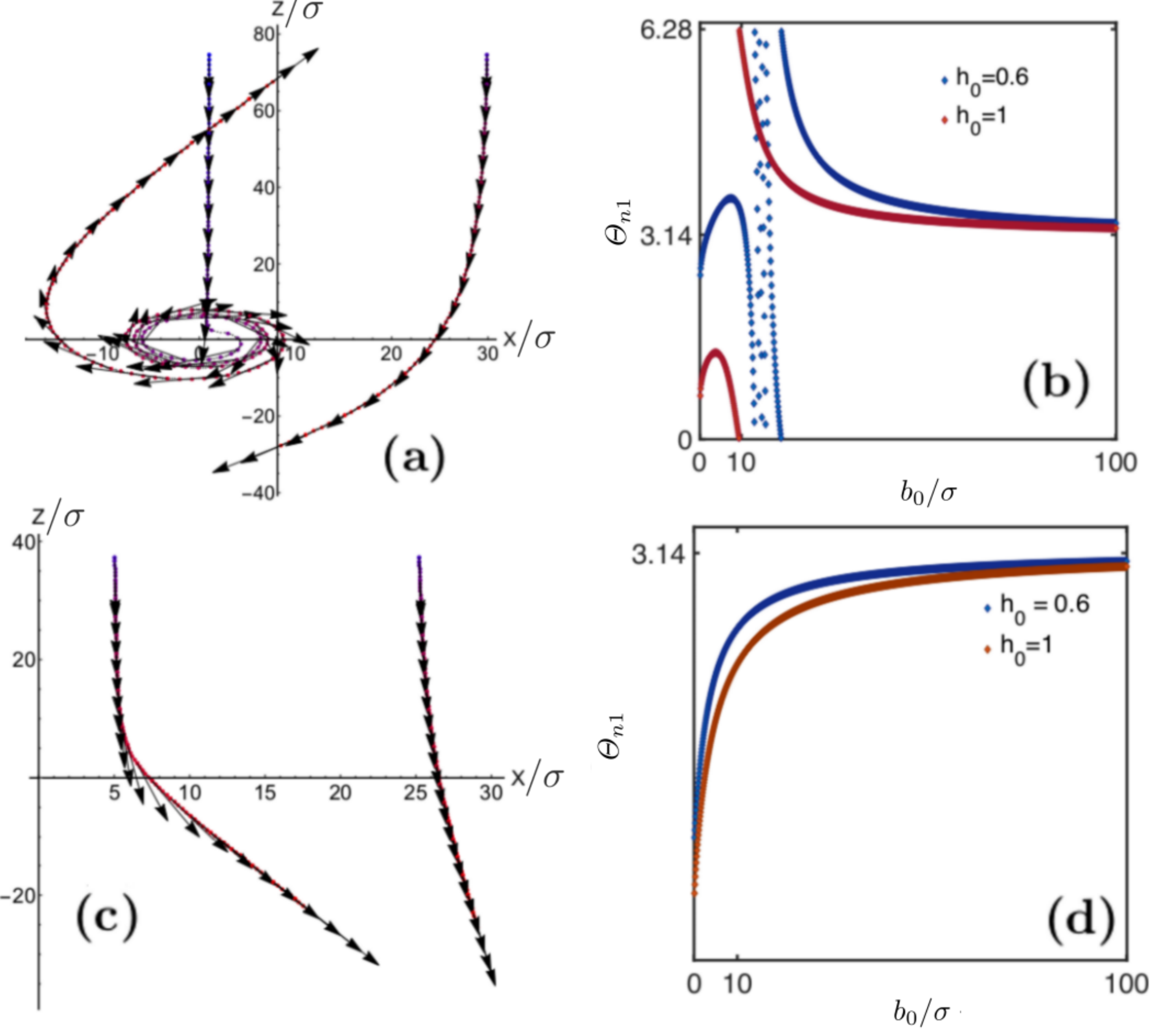}
		    \caption{Typical paths traced by chemotactic and anti-chemotactic swimmers when scattered by a product source placed at the origin are shown in figs. (a) and (c) respectively. The scattering angle $\Theta_{n1}$ varies with impact parameter $b_0$ as shown in figs. (b) and (d). $\Theta_{n1}$ are plotted as a function of the impact parameter $b_0$ for two values of $h_0$. The chemotactic swimmer revolves around the source multiple times before getting scattered, which explains the non-monotonic dependence of $\Theta_{n1}$ on $b_0$ in fig. (b). The anti-chemotactic swimmer saturates to $\Theta_{n1} = \pi$. The mobility design for the chemotactic swimmer are the same as in fig. \ref{Orbits} while  it is $\{1, 0.4, 0\}$ for the anti-chemotactic swimmer.}
\label{Scattering}
        \end{center}
\end{figure}
For large enough $h_0$, a chemotactic swimmer interacts briefly with the source before escaping it following paths that resemble scattering off an attractive centre, see fig. \ref{Scattering} (a). The scattering angle defined as $\Theta_{n1} = \lim_{t \to \infty} \theta_{n1}(t)$ is calculated by varying $h_0$ and the impact parameter $b_0$. $b_0$ is the initial lateral separation between the swimmer and the source. For a chemotactic swimmer at fixed $h_0$, there exists a threshold value of $b_0$ below which the swimmer is trapped in one of the two bound states.  Above a maximum $h_0$, the swimmer is scattered for all $b_0$. Fig. \ref{Scattering} (b) shows $\Theta_{n1}$ for two different values of $h_0$. A swimmer that can form orbits revolves around the source several times before escaping which results in non-monotonic dependence of the scattering angle $\Theta_{n1}$ on $b_0$. An anti-chemotactic swimmer constantly turns away from the source and gets repelled, see fig. \ref{Scattering} (c), (d).

\newpage
\subsection{Fixed points in an isotropic source} \label{SFixedPoint}
\begin{figure}[h]
\begin{center}
                  \includegraphics[angle=0,width=4.2in]{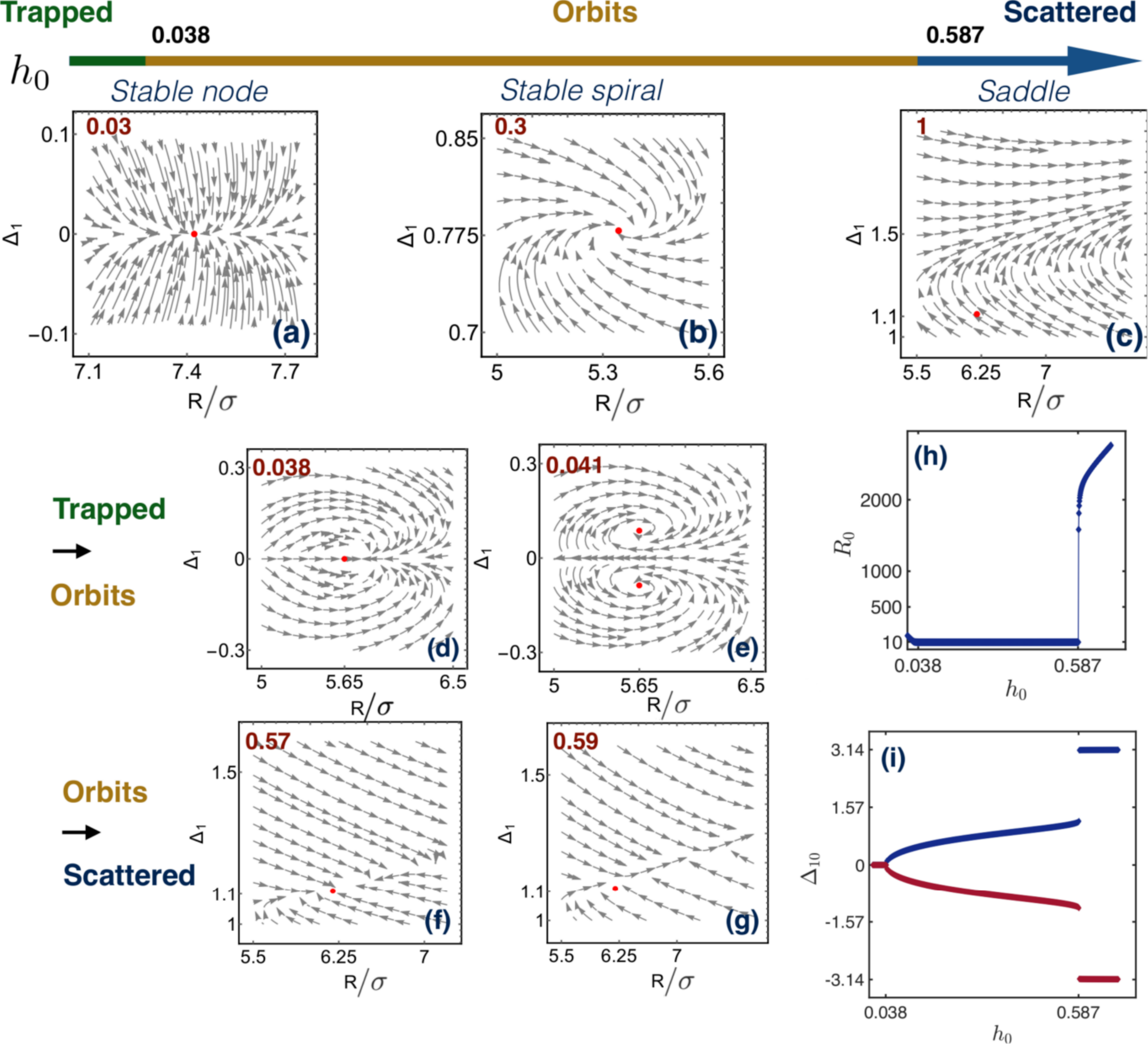}
		    \caption{{Phase portrait for a swimmer in an isotropic source showing typical trajectories with initial points close to the fixed point in the space spanned by $R$ and $\Delta_{1}$. The fixed points for the trapped state is $R_0>0, \,\, \Delta_{10} =0$, for orbits it is $R_0>0, \,\, \Delta_1 = \pm \Delta_{10}$ and for $\Delta_1 \approx \Theta_{n1}, \,\,  R \to \infty$.   Figures (a)-(c) show the flow lines for trapped, orbiting and scattered states respectively for the swimmer parameters same as in fig. \ref{Isotropic} (a); the value of $h_0$ is written on the top left. Figures (d)-(g) show the flow lines below and above the bifurcation points. Figures (h) and (i) show the variation of the steady state values of $\Delta_{10}$ and $R_0$ when $h_0$ is changed.}}
\label{FixedPoint}
        \end{center}
\end{figure}

To identify the two fixed points discussed so far, we will calculate the flow into these points (see fig. \ref{FixedPoint}). We consider an isotropic source only, as the fixed points and phase portrait in this case can be exactly represented in 2D space spanned by $R$ and $\Delta_1$. Close to fixed point $(R_0,\Delta_{10})$, the linearised equations can be written as $(\dot{\delta R}, \dot{\delta \Delta_1}) = \mathcal{M} \cdot (\delta R, \, \delta \Delta_1)$. The signs of the real parts of the two eigenvalues of the dynamical matrix $\mathcal{M}$ determine system stability. At low $h_0$ the fixed point is stable node, see fig. \ref{FixedPoint} (a), characterised by two unequal negative real eigenvalues. With increasing $h_{0}$ the system undergoes a pitchfork bifurcation to a stable spiral, see fig. \ref{FixedPoint} (b), i.e. the two eigenvalues are complex conjugate of each other with negative real parts. On increasing $h_0$ further, the system undergoes a spiral-saddle bifurcation as the real part of the eigenvalues, with the eigenvector predominantly along $R$, turns positive causing $R$ to grow unbounded, see fig. \ref{FixedPoint} (h). At this value of $h_0$ transition for a bound state to a scattering. The other eigenvalue remains negative and $\Delta_1$ reaches the fixed value $\pi$ as seen in fig. \ref{FixedPoint} (i). 

\newpage
\section{Two mobile swimmers}
The motion of one swimmer in the body fixed frame of another, studied in section \ref{SingleColloid} and the features discussed can be used to study two mobile swimmers. We show that the final state is independent of the microscopic details and can be sorted into three broad classes: (1) Active dimers, whose relative distance stays fixed while the centre of mass translates uniformly, (2) Binary swimmers, whose polar axes become phase-locked, so that they revolve around a common centre in synchronised closed circular orbits while maintaining a fixed distance between their centres, like planetary orbits under gravity. (3) Scattering states, where the swimmers interact for a finite period of time and then take off at an angle to one another.

The interactions between the swimmers are not reciprocal. As a result they can be mutually attractive, mutually repulsive or show a mixed response; when one of the two is attracted to the other, while the other is repelled. The translational velocities of the swimmers 1 and 2 are $ \bV_{1} = V(\Delta_1,\Delta_2,R) \hat{\bR} + \om(\Delta_1,\Delta_2,R) R \hat{\bm{\beta}}, \,\, \bV_{2} = -V(\Delta_2,\Delta_1,R) \hat{\bR} + \om(\Delta_2,\Delta_1,R) R \hat{\bm{\beta}}$ and the angular velocities are $\dot{\theta}_{n1}=\omega(\Delta_1,\Delta_2,R) , \,\, \dot{\theta}_{n2} = \omega(\Delta_2,\Delta_1,R)
$.  $R = |\bR_1 - \bR_2|$ is the separation between the two swimmers and $\hat{\bR}$ is the unit vector pointing from swimmer 2 to 1. The trajectories shown in the rest of this section are produced by solving the equations of motion with initial conditions $\tno(0)-\tnt(0) = \pi$, $X_{1}(0)-X_2(0) = b_0$, where $b_0$ is the impact parameter, and $Z_{1}(0)-Z_2(0)=10^3 \sigma $. We now discuss the three different types of dynamics in detail. 

\subsection{Active dimers} \label{ActiveDimer}
\begin{figure}[h]
\begin{center}
                  \includegraphics[angle=0,width=4.2in]{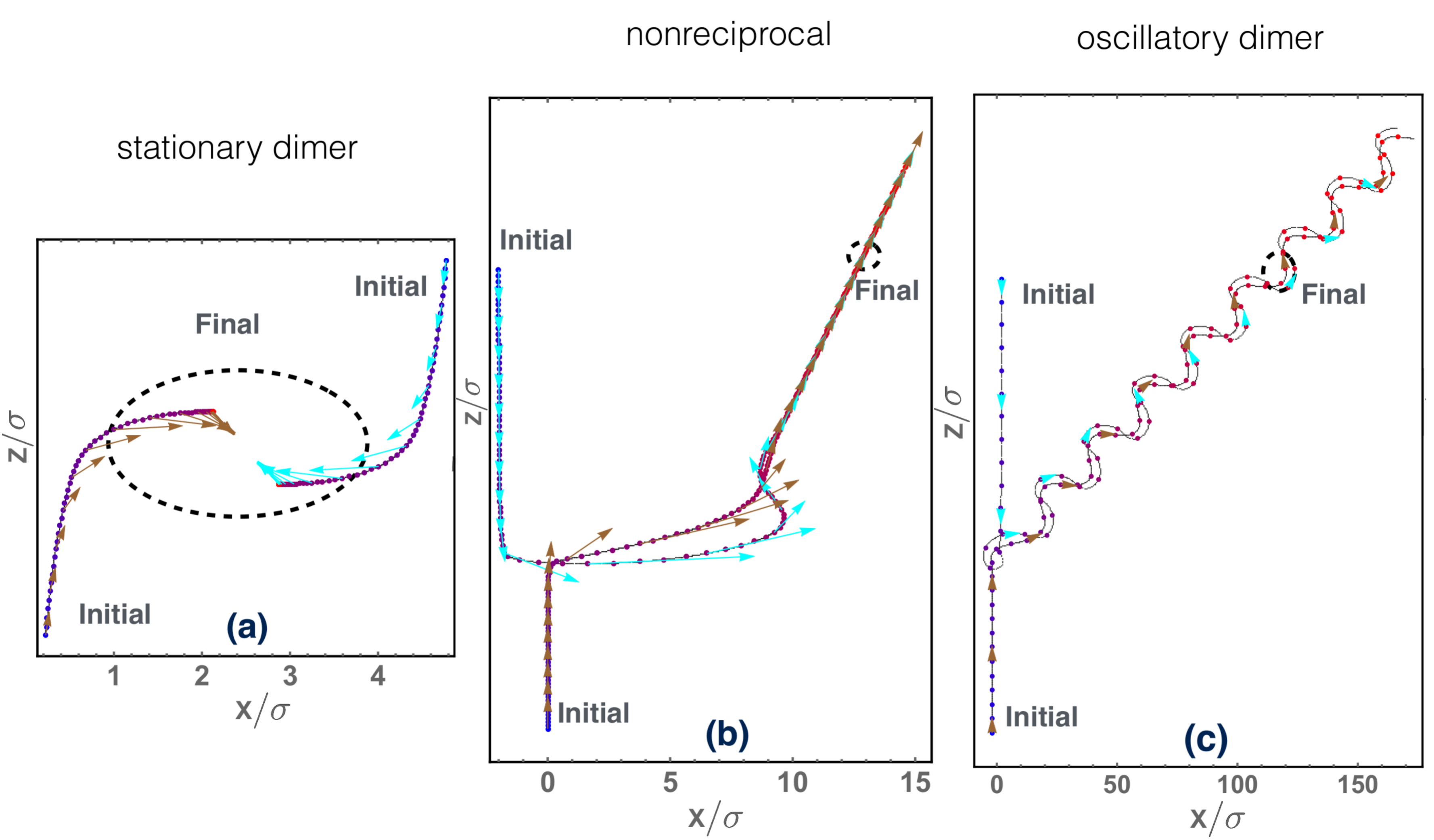}
	            \caption{{Active dimers are formed when at least one of the swimmers in the pair acts as a motile confining source for the other. Markers showing swimmer positions change from blue to red with time; polar axes of swimmers 1 and 2 are represented using cyan and brown arrows respectively.   In fig. (a), a pair of identical, mutually attractive swimmers forming a non-translating dimer. The catalytic and mobility coat designs are $\{1,-0.25,0\}$ and $\{1, -1.6, 0 \} $ respectively. In fig. (b) swimmer 1 chases swimmer 2 and  reach a final state when their polar axes both point in the direction of motion of the active dimer. The catalytic coat designs are $\{0.5, -0.5,0 \}$ and $\{1,-0.3,0 \}$ for 1 and 2 respectively. The mobility coat design is $\{1, -1.6,0 \}$ for 1 and $\{1,1.6,0 \}$ for 2 respectively. Figs. (c) shows looping trajectories, where $m^{(1)}=4 m^{(2)}$. Their catalytic coats are $\{1, -0.3,0 \}$ and $\{1.25,-0.5,0 \}$. Mobility coats are $\{1, -1.6,-2 \}$ and $\{1,1.6,-1.5 \}$ for 1 and 2 respectively.}}
\label{ActiveDimers}
        \end{center}
\end{figure}
Two mutually attractive or mixed swimmers form a stable bound state where their relative separation freezes at a constant value $R_0$ and their polar axes align. We call this final state which can be non-moving or self-propelled, an active dimer. This state is seen when one of the two swimmers is designed so that it can be trapped by a source. Two mutually attractive swimmers form an active dimer when their polar axes point inwards along the line joining their centres and their relative velocity vanishes, so that $R_0$ satisfies: $V_1(0,0,R_0) + V_2(0,0,R_0) = 0$. The dimer translates with the centre of mass velocity, defined in this inertia-less regime as the sum of the individual velocities as is given by $V_1(0,0,R_0) - V_2(0,0,R_0) $. A mixed pair also forms a dimer in a similar way, with the difference that one of them points towards the other while the other points away. The separation $R_0$ satisfies $V_1(0,0,R_0) + V_2(0,\pi,R_0) = 0$ and the translation velocity is $V_1(0,0,R_0) - V_2(0,\pi,R_0) $, where 1 is attracted to 2, but 2 is repelled by 1. Bound states due to forces are formed when a particle sits in the minima of the effective potential produced by the other. In this system, long-ranged attractive and short-ranged repulsive drift velocities are generated by a balance between attraction due to chemotaxis and repulsion due to phoresis. This can be contrasted with other propelling clusters where the symmetry is broken by the shape of the cluster \cite{Soto,DimerNonChem,Lowen_clusters}. For two identical attractive swimmers, the condition to form a stationary dimer is $V(0,0,R_0)  = 0$. The centre of mass of this system remains static always while the swimmers forms a stable dimer as `forces' balance.   

\subsection{Binary swimmers} \label{BinarySwimmers}
\begin{center}
\begin{figure}
      \hspace{0.5in}               \includegraphics[angle=0,width=3.7in]{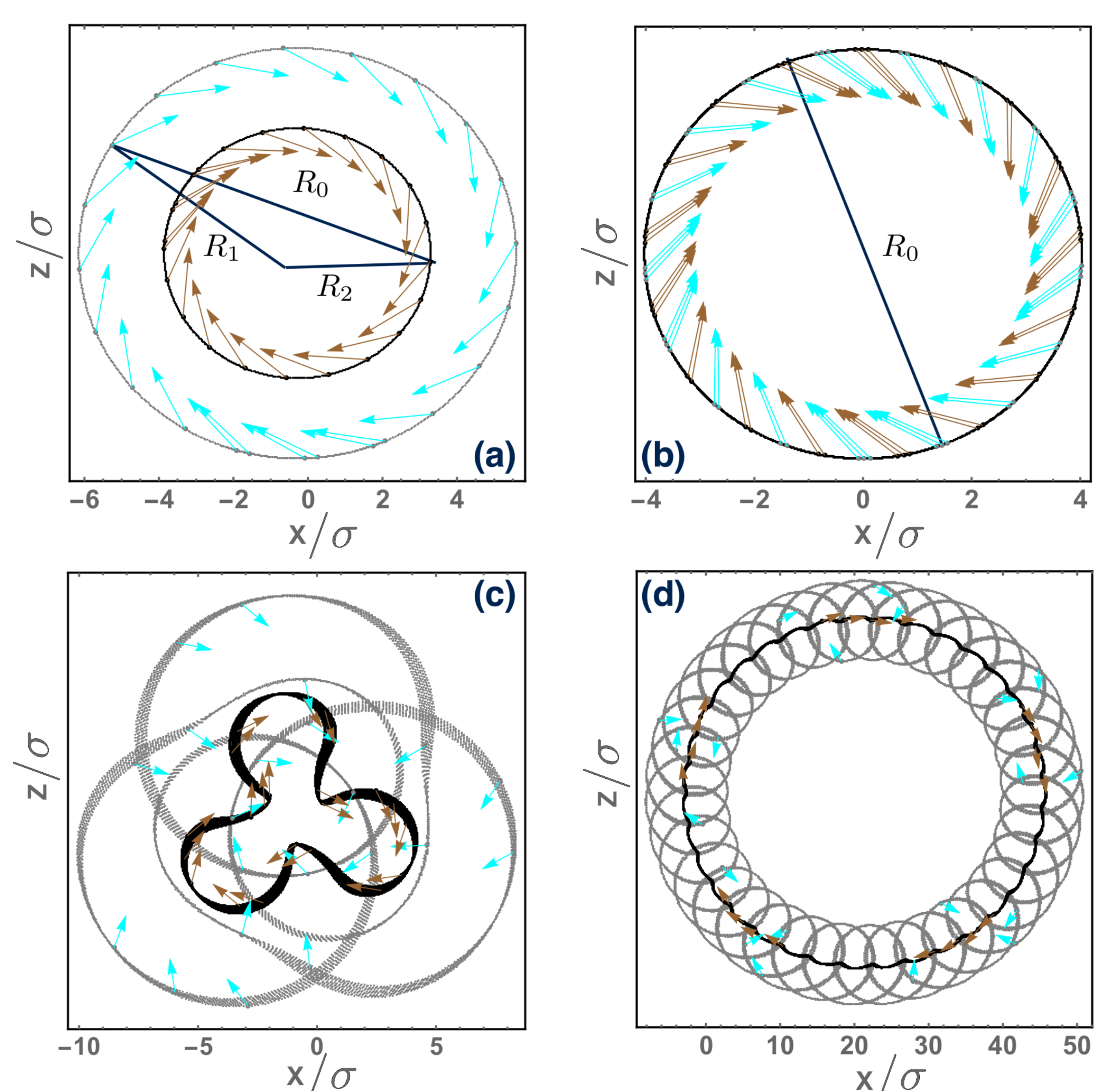}
          \caption{{Trajectories showing binary-swimmers that rotate in synchronised orbits around a common point. Polar axes of swimmers 1 and 2 are represented using cyan and brown arrows respectively; paths followed by 1 and 2 are shown in grey and black points. Fig. (a) shows two mutually attractive swimmers bound in a binary pair, fig. (b) shows a pair of identical swimmers rotating with $R_0$ equal to orbit diameter. Figs. (c) and (d) shows looping trajectories obtained when mobility of swimmer 1 is an order of magnitude larger than swimmer 2. The mobility of coat of all the swimmers in the figure are same as ones in Figure 5(a). The catalytic coat parameters are $\{1,-0.5,0\}$ and $\{3,-0.25,0\}$ for 1 and 2 in fig. (a), $\{1,-0.125,0\}$ in figs. (b) and (d) for both 1 and 2, $\{1,-0.15,0\}$ and $\{2,-0.25,0\}$ for 1 and 2 in fig. (c). }}
\label{Binary}
\end{figure}
\end{center} 
Two swimmers can go around a common point in synchronised circular orbits while maintaining fixed orientations $\Delta_1 = \Delta_{10}$ and $\Delta_2 = \Delta_{20}$, and a constant distance $R_0$ between their centres when at least one of them is designed such that it can orbit around a stationary source. We call this bound pair a binary swimmer for obvious reasons. The conditions of fixed relative distance and orientation are satisfied when
\beq
\fl
&& V(\Delta_{10},\Delta_{20},R_0) + V(\Delta_{20},\Delta_{10},R_0) = 0,\nonumber \\
\fl
 && \om(\Delta_{10},\Delta_{20},R_0) - \omega(\Delta_{10},\Delta_{20},R_0)= \om(\Delta_{20},\Delta_{10},R_0) - \omega(\Delta_{20},\Delta_{10},R_0)=0.  
 \label{sync2}
\eeq
The swimmers maintain fixed distances $R_{1,2} $ from their common centre. The total centre of mass velocity of the swimmer should be perpendicular to the line joining their centres to the common centre. This constraint and the equal orbital velocities of the swimmer provide three other conditions that are required to fully describe the bound state. We have the following relations for angles $\phi_{i0}$ between the sides of length $R_0$ and $R_i$
\beq
\fl
\cot \phi_{10} = \frac{V(\Delta_{10},\Delta_{20},R_0)}{R_0 \Omega(\Delta_{10},\Delta_{20},R_0)}, \, \, \cot \phi_{20} = \frac{V(\Delta_{20},\Delta_{10},R_0)}{R_0 \Omega(\Delta_{20},\Delta_{10},R_0)}.
\label{sync3}
\eeq
Eqs. \ref{sync2} and \ref{sync3} can be solved in order to obtain the stationary angles $\Delta_{10}$, $\Delta_{20}$, $R_0$ and $R_{1,2}$. The swimmers orbit around a common centre with in circular orbits with angular frequency $\omega(\Delta_{10},\Delta_{20},R_0)$. Fig. \ref{Binary} shows cases where synchronous orbits are possible. For swimmer mobilities $m^{(1)}>m^{(2)}$, the frequency of orbital motion are different obtain loopy trajectories as seen in Figure \ref{Binary} (c), (d).

For two identical swimmers, symmetry dictates that $\Delta_{10} = \Delta_{20}=\pi/2$. Thus identical binary swimmers are exactly like binary stars, as the component of their velocities parallel to the line joining the centre leads is zero and the non-zero perpendicular component leads to the rotation of the swimmers about a common point, see Figure \ref{Binary} (b). 

\subsection{Effect of fluctuations on bound states}
\begin{figure}[h]
\begin{center}
                  \includegraphics[angle=0,width=3.7in]{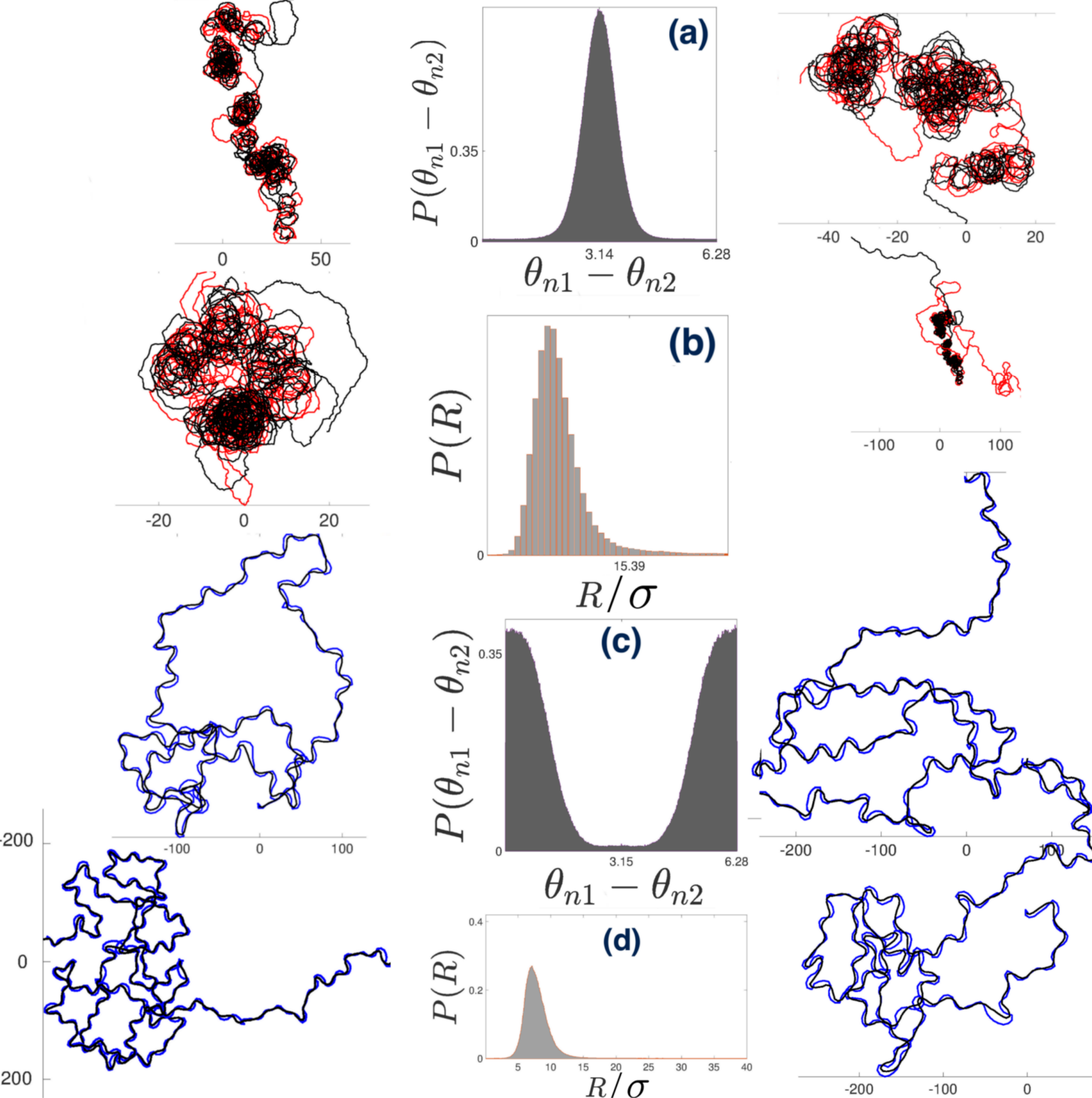}
    \caption{{Figure showing the stability of the two swimmer bound state to fluctuations. We add additive Gaussian noise to the equation of motion of two swimmers and solve the resulting stochastic dynamics assuming Ito statistics for the fluctuations. The trejectories in red and black are for the identical swimmers in fig. \ref{Binary} (b) and those in blue and black are for those in fig. \ref{Scat2} (b). The Peclet number for the swimmers is 60.  }}
\label{Noise2}
        \end{center}
\end{figure}

We end the discussion of bound states by testing the stability of these states to fluctuations. In order to do so additive brownian fluctuations are added to the equations of motion and distributions of $R$ and $\theta_{ni}$ are calculated in fig. \ref{Noise2}. The distrbutions peak at vaues $R_0$ and $\Delta_{10}+\Delta_{20}$ as calulated for the noise-free dynamics.

\subsection{Scattering}
\begin{figure}[h]
\begin{center}
                  \includegraphics[angle=0,width=4.3in]{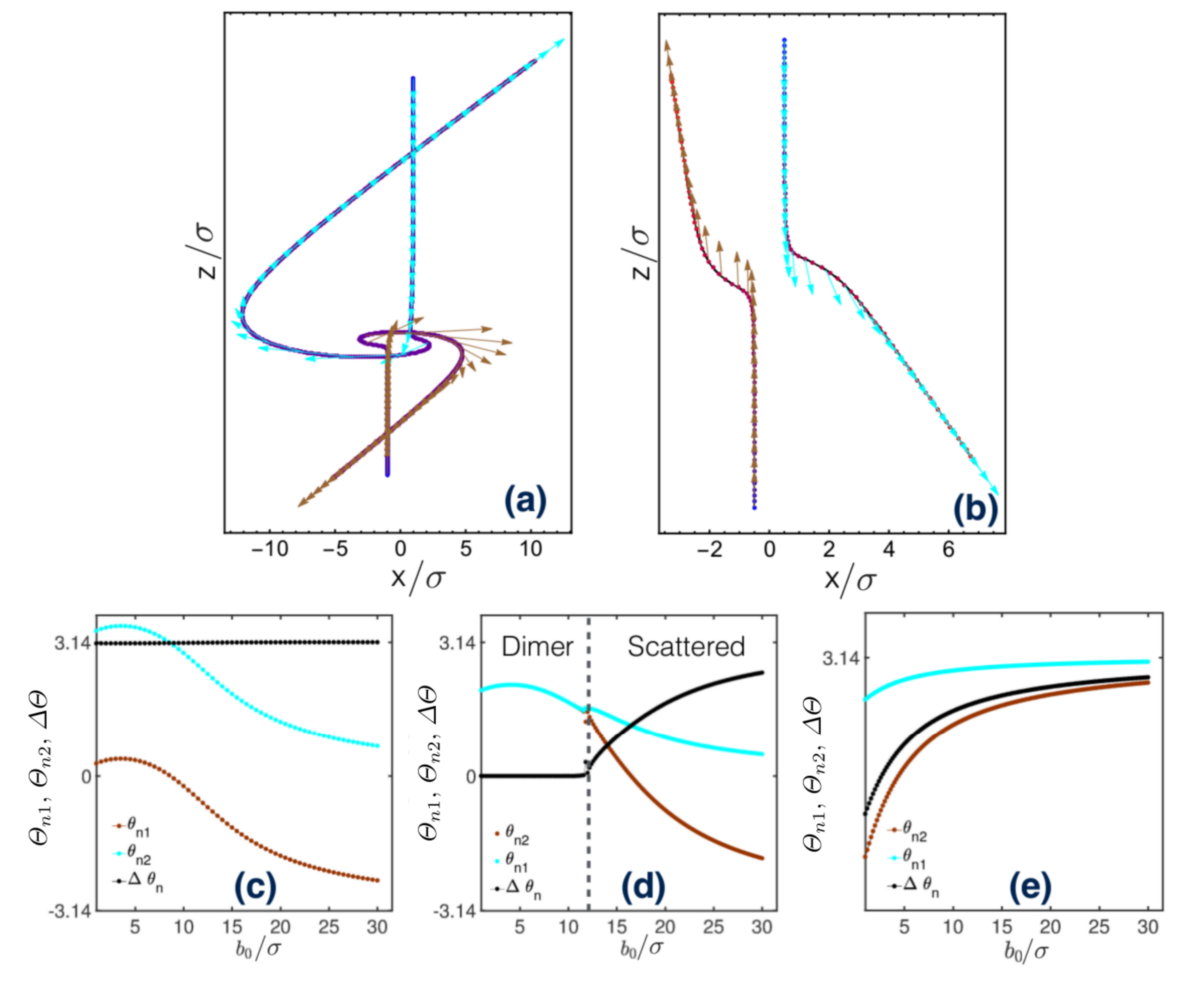}
		    \caption{{Two identical swimmers scatter off one another when self propulsion is stronger than the attraction. Fig. (a) shows typical trajectories of two mutually chemotactic swimmers and fig. (b) shows two mutually anti-hemotactic swimmers. The final polarity $\Theta_{n1}$, $\Theta_{n2}$ and the relative scattering angle $\Delta \Theta$ are shown as a function of the impact parameter for three cases: (c) two swimmers that scatter apart for all $b_0$, (d) two swimmers that form dimers for $b_0/\sigma < 13$, (e) two mutually anti-chemotactic swimmers.      }}
\label{Scat2}
        \end{center}
\end{figure}
Two mutually attractive or mixed swimmers scatter apart if the combination of self-propulsion and interaction along their centres is enough to separate them, see \ref{Scat2} (a). Two anti-chemotactic swimmers continously anti-align and scatter away, see \ref{Scat2} (b). The active dimers and binary-swimmers discussed in the subsections \ref{ActiveDimer} and \ref{BinarySwimmers} unbind for a large enough $b_0$, fig. \ref{Scat2} (d). The region of the state diagram in \ref{StateDiagram} where the swimmers scatter apart following trajectories typical of attractive or repulsive interactions are shown in white and blue respectively in fig. \ref{StateDiagram}.

\section{Conclusions}
We have explored theoretically the varieties of dynamics exhibited by polar self-diffusiophoretic
colloids, focusing on the case where the motion is planar. Gradients in the diffusing
chemical species -- generated by other colloids -- affect the
motion of the colloids, directly by translating them, and indirectly by rotating their
polarity and thus their self-propelling velocities. These effects cooperate and
compete, with remarkable dynamical consequences ranging from trapping, scattering and
simple orbits around a fixed source to complex pairs of dances, which are the main
results presented in this paper. The interactions are mediated by a diffusing field
and are therefore naturally long-ranged. A distinguishing feature of our treatment is
that the short-range interparticle repulsion is also dynamically generated by
phoretic mechanisms. The orientation vector carried by each particle sets the
direction of persistent motion. This endows the colloids with a kind of inertia. This is why, despite their vanishingly small Reynolds number, their behaviour bears some
similarity to scattering in Newtonian mechanics. The analogy is incomplete as there
is no equivalent of Newton's III Law. It would appear therefore that there is no
reduction of effective degrees of freedom in the 2-body problem; could it then show
chaos? We do not see evidence for this in our numerical calculations. Possibly this
is because the relaxational dynamics partially slaves some of the orientational
degrees of freedom to each other and to the separation vectors; see eq. \ref{expom} and eq. \ref{expvelang}. 
The case where the relaxation of chemical field occurs on
timescales comparable to that of swimmer orientations, or or when the finite size of
the swimmer creates orbits or active molecules \cite{Vasily_chemotaxis,lowen_loopy},
would be an interesting extension, as would the case of non-planar, i.e.,
three-dimensional, motion. Meanwhile, our predictions for the planar case can be
tested in suitable microfluidic setups \cite{ScatGeometry}.

A comparison to the Newtonian mechanics of particles interacting via central force
fields is natural: the concentration of diffusing species plays a role similar to a
potential, and the polar orientation resembles a momentum, endowing the dynamics with
a character similar to inertia as it determines the direction of persistent motion (see
\cite{DiscGravity} for another Stokesian driven system with an effective inertia).
The absence of a conserved centre-of-mass momentum, however, renders the analogy
incomplete, giving rise to the nontrivial joint motions of dimers discussed in
section IV. 

Hydrodynamics underlies the self-phoretic motility of our swimmers, but has not
otherwise been included in our treatment. The absence of the hydrodynamic interaction in our treatment, even in the residual form of an incompressibility constraint, means the appropriate experimental test of our predictions should be under planar confinement between walls that are permeable to solvent and ions. In light of recent work
\cite{MehranaStability} in which chemotaxis has been shown to compete against
instabilities originating due to fluid flows, it is of importance to examine
how the hydrodynamic interaction modifies the dynamics, especially of bound states,
that we have described in this work. Whereas we have concentrated on the two-body
problem, the implications of non-mutual pair interactions between dissimilar
particles for the collective behaviour of active binary mixtures remain to be
explored.  

We expect our work to inspire efforts to fabricate particles with a range of
catalytic and mobility coat patterns, using shape \cite{Shape1,Shape2} as an
additional control parameter, This will allow a test of our predictions through
exploration of our dynamical state diagrams. 

\def\ack{\ifletter\bigskip\noindent\ignorespaces\else
    \section{Acknowledgments}\fi}
SS thanks A Maitra, S Hablani, S Chatterjee, and K R Prathyusha for numerous
discussions. SS thanks A Maitra particularly for discussions on fixed points and
S Hablani for implementation of noisy dynamics. SS thanks R Seyboldt, A Maitra and T Adeleke-Larodo for detailed reading of the manuscript and offering many suggestions for improvement. SR was supported by a J C Bose Fellowship of the SERB, India, and a Homi Bhabha Chair Professorship of the Tata Education and Development Trust. 

\section{Appendix}
The appendix is organised in three sections: in (A) we present details of the parameters chosen to construct the state diagrams in fig. 2, in (B) we outline the steps leading to the equations of motion in eqs. \ref{expom}, \ref{expvelR} and \ref{expvelang} and in (C) the full equations of motion are written down.
\subsection{Appendix A: Parameters for the state diagram}
The state diagram is constructed by evaluating the trajectories for swimmers of different designs until a steady state is reached. Initial conditions are unchanged in evaluating a particular state diagram. Initial conditions for the single swimmer state diagrams in \ref{StateDiagram} (a), (b) and (c) are as follows:  $\theta_{n1}(0) = \pi$ and $Z_10)= 10^3 \sigma$, while $b_0 = 0.5 \sigma$ for figs. \ref{StateDiagram} (a) and \ref{StateDiagram} (b) and $b_0 = 2\sigma$ for fig. \ref{StateDiagram} (c). Design parameters for single swimmer state diagrams: in fig. \ref{StateDiagram} (a) The source is isotropic with catalytic coat $\{1,0,0\}$.  The swimmer catalytic coat design parameters are $\{1,-1,0\}$ and mobility $\mu_{0}^{(1)}=0.3$.  $\mu_1^{(1)}$, $\mu_2^{(1)}$ are varied in the range $(-0.4,0.25)$ and $(-1,1)$ in steps of $0.025$. In fig. \ref{StateDiagram} (b) we use swimmer parameters same as in \ref{StateDiagram} (a), in presence of an anisotropic source with catalytic coat $\{1,1,0\}$. $\mu_1^{(1)}$, $\mu_2^{(1)}$ are varied in the range $(-0.4,0.25)$ and $(0.6,1.2)$ in steps of $0.01$. In fig. \ref{StateDiagram} (c) The swimmer and source design are the same as in fig. \ref{StateDiagram} (a); and $h_0=1$ is larger than the value $0.587$ for which we find bound states. The impact parameter is $b_0=2 \sigma$. Note that only fig. \ref{StateDiagram} (c) is sensitive to choice of $b_0$. This is because given the choice of $h_0$, the capture radius for the swimmer is finite. If the impact parameter is increased larger than $30 \sigma$, the bound state disappear completely. 

The trajectories for two swimmers are evaluated for $\theta_{n1} - \theta_{n2} = 2\pi$ and $Z_1 - Z_2=2 \times 10^3 \sigma$. Parameters for two swimmer state diagrams: in fig. \ref{StateDiagram} (d), we consider two identical swimmers with $\mu^{(i)}_{0}=0.3$, $\mu^{(i)}_{1}=-0.5$ and $\mu^{(i)}_{2}$ is varied. The total catalytic activity $a^{(i)}$ is varied, the catalytic coat design is $\{ 1,-0.1,0 \}$ and $b_0=2$. In 1(e) we consider two swimmers with mobility $\{0.3, -0.5,1\}$, catalytic coat $\{1, -1, 0 \}$, $b_0=4 \sigma$ and surface activities $a^{(1)}$ and $a^{(2)}$ are varied. in fig. \ref{StateDiagram} (f) The total surface activity and $\mu^{(i)}_2$ is varied with other parameters being the same as in fig. \ref{StateDiagram} (e). 
\subsection{Appendix B: Evaluation of the chemical field}
\begin{figure}[h]
\begin{center}
                  \includegraphics[angle=0,width=2in]{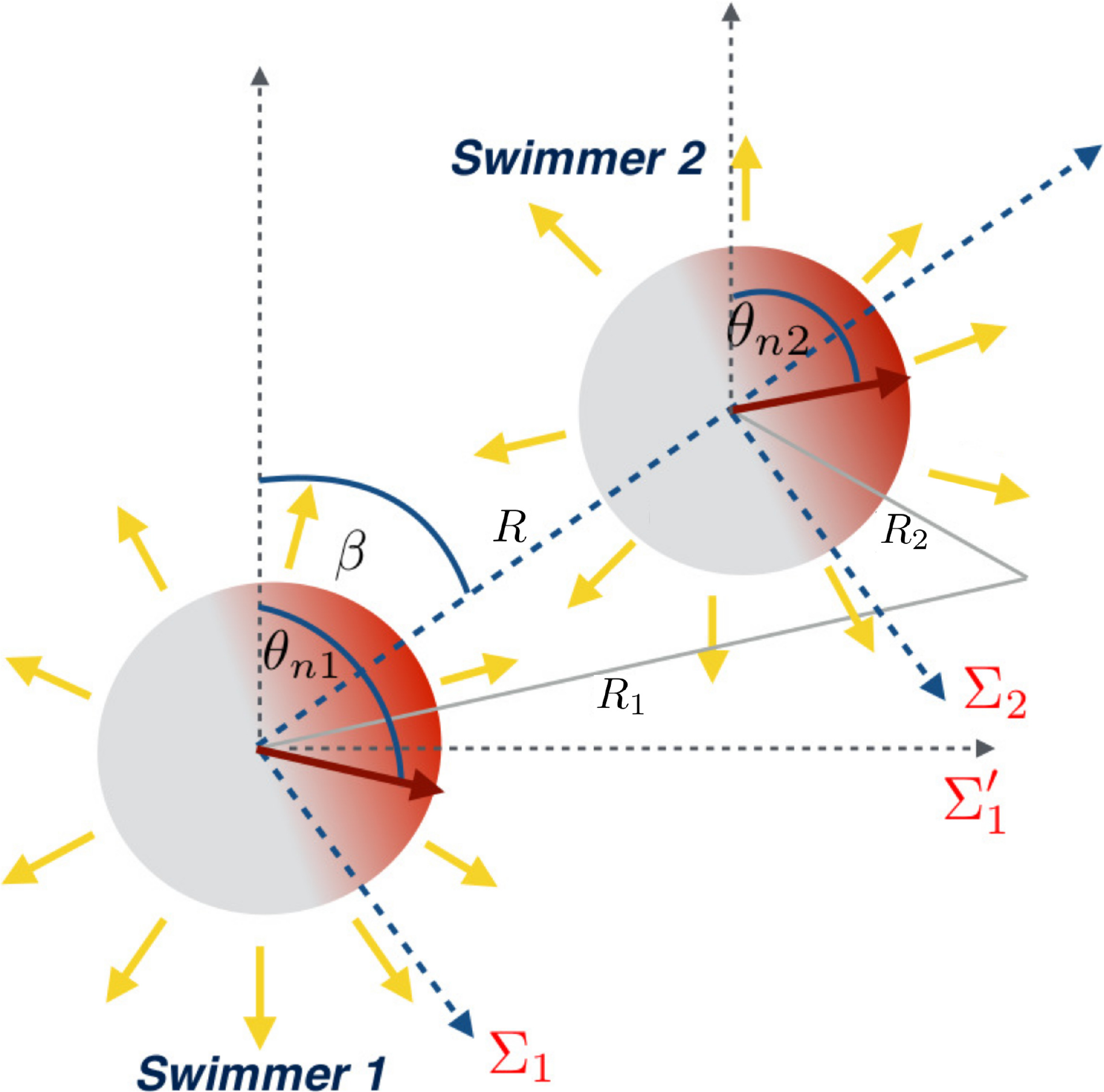}
		    \caption{{Figure showing two swimmers at a fixed distance and orientation and coordinate axes necessary for calculating the chemical field.}}
\label{2sphere}
        \end{center}
\end{figure}
In this section we outline a method of reflections to calculate the chemical field $\rho$ produced by two swimmers of any orientation with a separation $R$. $\rho$ satisfies Neumann boundary condition on the swimmer surfaces, (eq. \ref{TwoC} in the main text). 
\beq
\nabla^2 \rho  &=& 0, \label{rhoeq} \\  
-D\nabla_{\perp}\rho(\ba_i) &=& A^{(i)}(\ba_i) \label{bc}.
\label{eq1}
\eeq 
$\rho$ is a linear sum of contributions from both $A^{(i)}$, so it suffices to outline the calculation of $\rho$ for $a^{(1)}=0$. The solution can be split into two parts --  $\rho = {\rho}_{0}^{(2)} + \bar{\rho}$, where ${\rho}_{0}^{(2)}$ is the solution of the Poisson equation with the boundary condition eq. \ref{bc} on colloid 2 assuming colloid 1 to be absent. $\bar{\rho}$ is the correction due to presence of colloid 1 that we want to calculate. $\bar{\rho}$ is solved by a method of reflection, where at every order $n$, the boundary condition eq. \ref{bc} is enforced first on colloid 1 and then on 2. $\bar{\rho}$ is decomposed as follows
\beq
\bar{\rho} = \sum_{n} {\rho}^{(1)}_n + \sum_{n} {\rho}^{(2)}_n.
\label{expand}
\eeq
The convexity of the spherical surface rules out multiple reflections at a given order $n$.  ${\rho}^{(i)}_{n}$ satisfies the diffusion equation $\nabla^2 {\rho}^{(i)}_{n} = 0$ with boundary conditions
\beq
\fl
-\nabla_{\perp}  {\rho}^{(1)}_{n}(\ba_1) &=&  \sum_{k=0}^{n-1} \sum_{j=1}^{2} \nabla_{\perp} {\rho}^{(j)}_{k}(\ba_1), \label{bcmod1} \\  
\fl
-\nabla_{\perp}  {\rho}^{(2)}_{n}(\ba_2) &=&  \sum_{k=1}^{n'} \sum_{j=1}^{2} \nabla_{\perp} {\rho}^{(j)}_{k}(\ba_2),
\label{bcmod2}
\eeq
where $n' = n$ for $i=1$ and $n'=n-1$ for $i=2$. Also note that $\rho_{0}^{(1)}=0$. $\nabla_{\perp}  {\rho}^{(i)}_{n}(\ba_i)$ can be decomposed into surface harmonics as follows 
\beq
-\nabla_{\perp} {\rho}^{(i)}_{n}(\ba_i) \equiv \sum_{\ell m} A^{(i)}_{n,\ell m} Y_{\ell m}(\ba_i).
\label{EFexp}
\eeq
Substituting eq. \ref{EFexp} in eqs. \ref{bcmod1} and \ref{bcmod2}, we derive the following recursion relation for coefficients $A^{(1)}_{n,\ell m}$.  
\beq
A_{n,\ell m}^{(1)} =  \sum_{k=0}^{n-1} \sum_{j=1}^2  A^{(2)}_{k,jm} N^{12}_{\ell j m}.
\label{recursion}
\eeq
where $N^{12}_{j \ell m}$ is a geometrical factor that can be calculated as follows.

Recall from the main text, that the swimmers have radii $\sigma$, polar axes $\theta_{ni}$. In fig. \ref{2sphere}, the frames of reference $\sum_1$ and $\sum_2$  are in the body fixed axes of the colloids and their $z$ axes are parallel to the line joining their centres. $\sum_1'$ is the lab frame with its origin coinciding with that of $\sum_1$. We use spherical coordinates $(R_1,\theta_1,\phi_1)$ and $(R_2,\theta_2,\phi_2)$ in frames $\sum_1$ and $\sum_2$ respectively. The unit vectors representing polar axes are $(1,\theta_{n1},0)$ and $(1,\theta_{n2},0)$, see fig. \ref{2sphere}. The expression for $\rho^{(1)}_{0}$ satisfying bounday conditions eq. \ref{EFexp} is
\begin{eqnarray}
\rho^{(2)}_{n} &=& {\sum_{\ell m}}  \frac{1}{(\ell+1)R_2^{\ell+1}} {Y_{\ell m}(\theta,\phi)A^{(2)}_{n,\ell m}}
\label{pb1}
\end{eqnarray}
The following transformations connect coordinates in $\sum_{1}$ and  $\sum_{2}$
\begin{eqnarray}
\theta_2(\theta_1,R_1,R) &=& \cos^{-1}\left(\frac{R_1 \cos \theta_1 - R}{\sqrt{R^2+R_1^2-2 R_1 R \cos \theta_1}}\right) \nonumber \\
R_2(\theta_1,R_1,R) &=& \sqrt{R^2+R_1^{2}-2 R_1 R \cos \theta_1} \nonumber \\
\phi_2 &=& \phi_1 
\label{trans}
\end{eqnarray}
Using eq. \ref{trans} in eq. \ref{pb1}, we obtain $\rho_{0}^{(1)}$ in $\sum_1$. Define
\begin{eqnarray}
&& N^{12}_{i\ell m} = \sigma \partial_{R_1} \int Y^{*}_{\ell m}(\theta,\phi) \frac{  Y_{im} \left[ \theta_2(\theta_1,R_1,R), \phi\right]} {R_2(\theta_1,R_1,R)^{\ell+1}} \mbox{d} \ba_1.
\label{coeff}
\end{eqnarray} 
A similar set of coefficients $N^{21}_{i \ell m}$ are defined by exchanging indices $1$ and $2$ in eq. \ref{coeff}. The recursion relations eq. \ref{recursion} can be solved to obtain the field to any degree of precision. In order to obtain the equations of motions presented in the main text we truncate the expansion at $n=1$. For a uniformly coated colloid 2 with $\sigma^{(2)}_{0}=1$ we have the following result consistent with the potential due to a charge next to a sphere with Neumann boundary conditions (see \cite{JDJackson})
\beq
\fl
\rho_0 + \rho_1^{(1)} = \frac{a^{(2)} \sigma}{D} \sum_{\ell m} \frac{4 \pi}{2 \ell+1} \left[  \frac{\sigma R_1^\ell}{R^{\ell+1}} + \frac{\ell }{\sigma(\ell+1)}  \left( \frac{\sigma^2}{R R_1  } \right)^{\ell+1}  \right] Y_{\ell m}(\theta,\phi) Y^*_{\ell m}(\beta,0).
\eeq

\subsection{Appendix C: complete equations of motion}
The angular momentum is
\beq
\fl
&& \omega(\Delta_1,\Delta_2,r  ) =  \frac{3 \sigma^2 \alpha_0^{(2)} \mu_1^{(1)}}{8 R^2}  \sin \Delta_1  +  \frac{3 \sigma^3 \alpha_0^{(2)} \mu_2^{(1)}}{8R^3} \sin 2 \Delta_1  - \frac{3 \sigma^3 \alpha_1^{(2)} \mu_1^{(1)}}{32R^3} \left[ \sin \Delta_1 - 3 \sin (\Delta_1 + \Delta_2)  \right] \nonumber \\
\fl
&&  -\frac{3 \sigma^4 \alpha_1^{(2)}  \mu_2^{(1)} }{32 R^4} \left[ 5 \sin (2 \Delta_1 + \Delta_2 ) - \sin(2 \Delta_1 - \Delta_2)  \right] + \frac{3 \sigma^4 \alpha_2^{(2)} \mu_1^{(1)} }{64 R^4} \left[ 5 \sin (\Delta_1 + 2\Delta_2) + 2 \sin \Delta_1 + \sin(\Delta_1 - 2 \Delta_2) \right]   \nonumber \\
\fl
&& +\frac{\sigma^5 \alpha_2^{(2)} \mu_2^{(1)}}{64 R^5} \left[  3 \sin (2\Delta_1 -2 \Delta_2) + 35 \sin(2\Delta_1 + 2\Delta_2) + 10 \sin 2 \Delta_1 \right]
\eeq
The radial velocity is
\beq
\fl
&& V(\Delta_1,\Delta_2,r  ) =  -\left( \mu_{0}^{(1)} - \frac{\mu_2^{(1)}}{20} \right) \frac{\sigma^2 \alpha_0^{(2)}}{R^2} + \frac{3 \sigma^2 \mu_2^{(1)} \alpha_0^{(2)} }{20 R^2} \cos 2 \Delta_1 , \nonumber \\ 
\fl
&& - \frac{2 \sigma^3 \alpha_0^{(2)} \mu_1^{(1)} }{3 R^3}  \cos \Delta_1  + \frac{  \sigma^3 \alpha_1^{(2)} \mu _{10}}{R^3} \cos \Delta_2   + \frac{\sigma^3 \mu_2^{(1)} \alpha_1^{(2)}}{80 R^3} \left[ 9 \cos (2 \Delta_1+ \Delta_2)+3 \cos (2 \Delta_1- \Delta_2)+4 \cos \Delta_2 \right] \nonumber \\ 
 \fl
&& +\frac{ \sigma^4 \alpha_1^{(2)} \mu_1^{(1)}}{4 R^4} [3 \cos (\Delta_1+\Delta_2)+\cos (\Delta_1-\Delta_2)]  - \frac{ \sigma^4 \alpha_2^{(2)} \mu _{10}}{4 R^4}  [1+3 \cos 2 \Delta_2]   \nonumber \\ 
\fl
&& + \frac{\sigma^4 \mu_2^{(1)} \alpha_2^{(2)}}{160 R^4} [6 \cos 2 \Delta_1 +15 \cos (2 \Delta_1+2 \Delta_2)+3 \cos (2 \Delta_1-2 \Delta_2)+6 \cos 2 \Delta_2+2] 
 \nonumber \\ 
\fl
&& -\frac{ \sigma^5 \mu_1^{(1)} \alpha_2^{(2)}}{6 R^5} [2 \cos \Delta_1 +5 \cos (\Delta_1 +2 \Delta_2)+\cos (\Delta_1 -2 \Delta_2)] .
\label{RadVel}
\eeq
The orbital angular velocity is
\beq
\fl
&& \Omega(\Delta_1,\Delta_2,r  ) = \frac{3 \sigma^3 \mu_2^{(1)} \alpha_0^{(2)}}{20 R^3} \sin 2 \Delta_1  + \frac{ \sigma^4 \mu_1^{(1)} \alpha_0^{(2)}}{3R^4} \sin \Delta_1 - \frac{\mu _{10} \alpha_1^{(2)}}{2 R^4} \sin  \Delta_2    \nonumber \\ 
\fl
&&  + \frac{\sigma^4 \mu_2^{(1)} \alpha_1^{(2)}}{80 R^4} [2 \sin \Delta_2 -9 \sin (2 \Delta_1+\Delta_2) - 3 \sin (2 \Delta_1-\Delta_2)] - \frac{\mu_1^{(1)} \alpha_1^{(2)}}{2R^5} \sin (\Delta_1+\Delta_2)  + \frac{\sigma^5 \mu _{10} \alpha_2^{(2)}}{2R^5} \sin 2 \Delta_2  \nonumber \\ 
\fl
&& + \frac{\sigma^5 \mu_2^{(1)} \alpha_2^{(2)}}{160 R^5} \left[ 6 \sin (2 \Delta_1)+15 \sin (2  \Delta_1+2  \Delta_2) +3 \sin (2  \Delta_1-2  \Delta_2)  - 4 \sin 2  \Delta_2 \right] \nonumber \\
\fl
&& + \frac{\sigma^6 \mu_1^{(1)} \alpha_2^{(2)}}{24 R^6} [2 \sin (\Delta_1)+15 \sin (\Delta_1+2 \Delta_2)-\sin (\Delta_1-2 \Delta_2)].
\label{AngVel}
\eeq

\end{document}